\pdfoutput=1
\documentclass[prd,aps,nofootinbib,twocolumn,superscriptaddress,preprintnumbers,balancelastpage,longbibliography]{revtex4-1}
\usepackage{graphicx}	
\usepackage{amsmath}	
\usepackage{dcolumn}
\usepackage{bm}
\usepackage{graphics}
\usepackage{afterpage}
\usepackage{float}
\usepackage{subfigure}
\usepackage{rotating}
\usepackage{multirow}
\usepackage{tabularx}
\usepackage{booktabs}
\usepackage{multirow}
\usepackage{fancyhdr}
\usepackage{hyperref}
\usepackage[utf8]{inputenc}
\usepackage{theorem}
\usepackage{moreverb}
\usepackage{euscript}
\usepackage{psfrag}
\usepackage{slashed}
\usepackage{mathtools}
\usepackage{makecell}
\usepackage[flushleft]{threeparttable}
\usepackage{adjustbox}

\usepackage{dcolumn}
\usepackage{bm}
\usepackage[mathlines]{lineno}
\usepackage{lettrine}
\usepackage[dvipsnames]{xcolor}
\usepackage{blindtext}
\usepackage{soul}

\input Zallman.fd

\LettrineTextFont{\itshape}
\hypersetup{
     colorlinks   = true,
     citecolor    = orange,
     urlcolor     = orange,
     linkcolor    = orange
}

\usepackage{listings}
\usepackage{color,xcolor}

\newcommand{\appropto}{\mathrel{\vcenter{
  \offinterlineskip\halign{\hfil$##$\cr
    \propto\cr\noalign{\kern2pt}\sim\cr\noalign{\kern-2pt}}}}}

\begin{document}
\preprint{SLAC-PUB-17771}

\title{Dark Kinetic Heating of Exoplanets and Brown Dwarfs}
\author{Javier F. Acevedo}
\thanks{ \href{mailto:jfacev@slac.stanford.edu}{jfacev@slac.stanford.edu}; \href{https://orcid.org/0000-0003-3666-0951}{0000-0003-3666-0951}}
\affiliation{Particle Theory Group, SLAC National Accelerator Laboratory, Stanford, CA 94035, USA}

\author{Rebecca K. Leane}
\thanks{\href{mailto:rleane@slac.stanford.edu}{rleane@slac.stanford.edu}; \href{http://orcid.org/0000-0002-1287-8780}{0000-0002-1287-8780}}
\affiliation{Particle Theory Group, SLAC National Accelerator Laboratory, Stanford, CA 94035, USA}
\affiliation{Kavli Institute for Particle Astrophysics and Cosmology, Stanford University, Stanford, CA 94035, USA}

\author{Aidan J. Reilly}
\thanks{ \href{mailto:areilly8@stanford.edu}{areilly8@stanford.edu}; \href{https://orcid.org/0009-0005-4810-8920}{0009-0005-4810-8920}}
\affiliation{Particle Theory Group, SLAC National Accelerator Laboratory, Stanford, CA 94035, USA}

\begin{abstract}
Dark kinetic heating of neutron stars has been previously studied as a promising dark matter detection avenue. Kinetic heating occurs when dark matter is sped up to relativistic speeds in the gravitational well of high-escape velocity objects, and deposits kinetic energy after becoming captured by the object, thereby increasing its temperature. We show that dark kinetic heating can be significant even in objects with low-escape velocities, such as exoplanets and brown dwarfs, increasing the discovery potential of such searches. This can occur if there is a long-range dark force, creating a ``dark escape velocity", leading to heating rates substantially larger than those expected from neutron stars. We consequently set constraints on dark sector parameters using Wide-field Infrared Survey Explorer and JWST data on Super-Jupiter WISE 0855-0714, and map out future sensitivity to the dark matter scattering cross section below $10^{-40}$~cm$^2$. We compare dark kinetic heating rates of other lower escape velocity objects such as the Earth, Sun, and white dwarfs, finding complementary kinetic heating signals are possible depending on particle physics parameters.
\end{abstract}

\maketitle

\lettrine{N}{eutron stars} have been previously hailed as one of the supreme astrophysical detectors for dark matter. Their high densities afford them relativistic escape velocities, leading to incoming dark matter sped-up to speeds two thirds that of light. If this high-speed dark matter becomes trapped, substantial kinetic energy is absorbed by the neutron star, increasing its temperature. This signal has been referred to as ``dark kinetic heating"~\cite{Baryakhtar:2017dbj}, and provides a signal for an expanded range of particle physics models, extending to not only weakly interacting massive particles (WIMPs), but also those such as asymmetric dark matter ($i.e.$ where dark matter does not annihilate), inelastic dark matter, or even pure Higgsinos~\cite{Baryakhtar:2017dbj}. No other star or planet has an escape velocity anywhere near that of neutron stars, such that dark kinetic heating was previously only thought detectable in these ultra-dense objects.

Dark matter heating of low-escape velocity objects has been considered, but only with dark matter annihilation heat rather than kinetic heating. One class of ideal, low-escape velocity dark matter detectors where dark matter annihilation heat has been considered includes exoplanets and brown dwarfs~\cite{Leane:2020wob}. These objects are ideal because they have lower core temperatures relative to $e.g.$ nuclear-burning stars, leading to retention of lighter dark matter, as well as lower heat backgrounds~\cite{Leane:2020wob,Leane:2024bvh}. Dark matter heating in exoplanets and brown dwarfs is also likely much easier to detect than neutron stars, due to their substantially larger surface areas and therefore larger potential luminosities. Exoplanets and brown dwarfs can therefore be detected far into the Galactic center, providing a potential tracer of the unknown dark matter density distribution in the Galaxy~\cite{Leane:2020wob,Benito:2024yki}. There are a large number of candidates detected or soon to be detected, with the onslaught of new high-power infrared and optical telescopes such as JWST, Roman, and Rubin, making this search promising~\cite{Leane:2020wob,Perryman_2014,green2012widefield,Johnson_2020,Leane:2024bvh,Benito:2024yki}.

\begin{figure}[t!]
    \centering
\includegraphics[width=0.92\columnwidth]{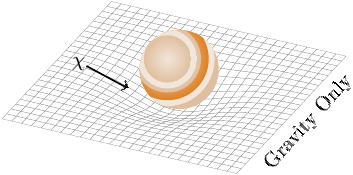}
\includegraphics[width=0.98\columnwidth]{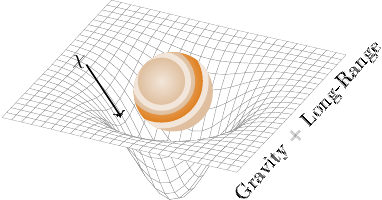}

    \caption{Schematic of dark kinetic heating for traditionally low-escape velocity objects. The addition of a new long-range force increases the escape velocity of objects, accelerating the dark matter particle $\chi$ to relativistic speeds and leading to large amounts of detectable dark kinetic heating.}
    \label{fig:schem}
\end{figure}

In this work, we point out that objects which are traditionally thought of as low-escape velocity objects can in fact yield substantial amounts of detectable kinetic heating. This can occur if there are long-range forces between dark matter particles and/or between dark matter and Standard Model particles, as the additional interaction can serve to increase the escape velocity of such objects. This is possible if the long-range interaction is attractive, which is generally expected for scalar interactions, and will occur for vector interactions if the charge assignments attract. We focus on long-range dark matter self-interactions, which generally drive this effect for asymmetric dark matter, creating a dark escape velocity. This scenario is shown schematically in Figure~\ref{fig:schem}. 

We will focus on example realizations of this effect for exoplanets and brown dwarfs, due to their many advantages described above. Increased amounts of dark kinetic heating from long-range forces were considered previously for neutron stars in Ref.~\cite{Gresham:2022biw}, where the escape velocity is already relativistic in the gravity-only case. We will show instead that low-escape velocity objects can afford dark kinetic heating rates even larger than that expected from neutron stars, depending on model parameters. Furthermore, as dark kinetic heating can be substantial even in the local position, rather than the usually required larger signal closer to the Galactic center, we will study existing infrared measurements of the coldest nearby Super-Jupiter with JWST and the Wide-field Infrared Survey Explorer (WISE), which can already set constraints. We will also examine future prospects to discover dark matter via dark kinetic heating throughout the Galaxy.

Our paper is organized as follows. In Sec.~\ref{sec: model set up}, we outline our toy model setup with long-range dark matter self-interactions. In Sec.~\ref{sec:capture}, we discuss capture rates under our long-range setup, and link these to the expected temperatures from dark kinetic heating in Sec.~\ref{sec:kinetic}. In Sec.~\ref{sec:bd}, we apply our setup to exoplanets and brown dwarfs to investigate the relationship between particle physics parameters of the long-range force, and demonstrate that existing infrared measurements of the Super-Jupiter WISE 0855-0714 allow constraints on models to already be set. We also detail future projections for detection of other Super-Jupiters or brown dwarfs, showing that dark kinetic heating allows for much lower scattering cross section sensitivity than that expected from dark matter annihilation heat alone. In Sec.~\ref{sec: other objects} we investigate the complementarity of the parameter space for a range of other low-escape velocity objects such as the Earth, Sun, and white dwarfs. We conclude in Sec.~\ref{sec:conclusion}.

\section{Particle Physics}
\label{sec: model set up}

\subsection{Toy Model Setup}

To demonstrate the dark kinetic heating effect in low-escape velocity objects, we will consider a simple scenario consisting of a light scalar mediator $\phi$ with mass $m_\phi$, that interacts with fermionic non-annihilating dark matter particles $\chi$ with a coupling strength of $g_\chi$,
\begin{equation}
\label{eq: lagrangian}
   \mathcal{L} \supset \frac{1}{2}(\partial_\mu \phi)^2 - \frac{1}{2}m_\phi^2 \phi^2 + \bar{\chi}\left(i \gamma^{\mu} \partial_\mu - m_\chi\right)\chi - g_\chi \phi \bar{\chi} \chi\,,
\end{equation} 
where $m_\chi$ is the dark matter mass. It is also of course possible to consider a Standard Model coupling to $\phi$, as was considered previously for neutron star kinetic heating~\cite{Gresham:2022biw}. However, experimental constraints on the Standard Model coupling typically lead to weaker signals for the same value of coupling $g_\chi$. On the other hand, $g_\chi$ is very weakly constrained, with the strongest constraints claimed by observations of the Bullet Cluster and halo shapes, which presently constrain $g_\chi \lesssim 10^{-4} \, (m_\chi/{\rm MeV})^{3/4}$ \cite{Knapen:2017xzo}. We therefore focus on long-range dark matter self-interactions. Lastly, higher-order interactions of $\phi$ are in principle allowed as well, but we will consider scenarios where such terms are negligible. We discuss extensions to a non-zero quartic interaction in Appendix~\ref{app: screening}. 

A general feature of a long-range attractive force is the presence of the Yukawa potential
\begin{equation}
    \label{eq:yukawa}
    V(r) = -\frac{\alpha}{r}e^{-r/\lambda}~,
\end{equation}
which can be sourced by our setup in Eq.~(\ref{eq: lagrangian}), where for our dark matter self-interaction scenario of interest, $\alpha=g_\chi^2/4\pi$, $\lambda=m_\phi^{-1}$ is the dark mediator force range, and $r$ is the separation distance. 

\subsection{Physical Effects and Motivations}

Once sufficient dark matter is captured by a celestial object, the potential in Eq.~(\ref{eq:yukawa}) introduces three primary effects on heating that distinguish the calculation from the standard short-range scenario. First, the long-range force enhances gravitational focusing, causing more dark matter to intersect the object than would occur through gravity alone. Second, dark matter arrives at the surface with increased kinetic energy due to the additional potential. Third, the dark matter may undergo self-scattering via the long-range interaction. In the presence of an additional sufficiently strong contact interaction between dark matter and the Standard Model, however, we can simplify the calculation by ignoring the third effect. For simplicity, we adopt this assumption and calculate conservative sensitivities by assuming a large enough cross section to facilitate capture. While some interaction with baryons is necessary for observable heating, it need not dominate the capture process. Our goal is to establish, for the first time, that detectable dark kinetic heating can occur in objects with low escape velocities. As such, we begin with this simplified case and leave a general treatment for future work.

Another possible physical effect is the formation of multi-particle bound states. However, for the couplings and interaction ranges considered here, such bound state formation is suppressed (see, $e.g.$, Ref.~\cite{Gresham:2017cvl}). Accordingly, our heating signal arises from the energy deposition of individual dark matter particles.  One may also worry that this long range force will introduce a new Jeans length for the Galactic dark matter halo. However, this does not occur for the range of couplings and mediator masses of the interaction we consider; see Appendix~\ref{app: Jeans} for further discussion.

Finally, we note the broader motivation for considering dark matter self-interactions. These are well-motivated on both observational and theoretical grounds. For example, the wide diversity of inferred dark matter halo profiles on small scales is in tension with predictions from collisionless cold dark matter simulations (see $e.g.$ Ref.~\cite{Tulin:2017ara} for a review). Self-interactions within the dark sector have been proposed as a resolution to this small-scale structure problem, while maintaining consistency with the successes of $\Lambda$CDM at large scales~\cite{Kaplinghat:2013xca,Elbert:2014bma,Robertson:2017mgj}. From a model-building perspective, long-range forces—mediated by either vector or scalar particles—remain viable. Scalar-mediated forces, in particular, are expected in scenarios such as those involving quintessence~\cite{Farrar:2003uw,PhysRevD.66.043528}. More recent studies have also explored collective effects of long-range forces in the dark sector~\cite{Bogorad:2023wzn}, though none to date have examined forces with the specific range considered in our work.

\section{Capture Framework}
\label{sec:capture}

\subsection{Dark Matter Inputs}
\label{sec:inputs}

To apply our formalism of dark kinetic heating to celestial objects, we first need physical dark matter parameter inputs. We take a Maxwell-Boltzmann distribution for the dark matter velocity profile,
\begin{equation}
    \begin{split}
    f(v_\chi) = \frac{4}{\sqrt{\pi}v_p^3} v_\chi^2 \, e^{-v_\chi^2/v_p^2} \,,
    \end{split}
\end{equation}
with an average velocity of $v_p = 270$ km/s~\cite{Lin_2019}, which is a conservative choice for regions towards the Galactic center.

Furthermore, we will consider three possible dark matter distributions which feature varying densities towards the Galactic center: a Navarro-Frenk-White (NFW) profile \cite{Navarro_1996}, a generalized NFW (gNFW) profile motivated by adiabatic contraction~\cite{2011arXiv1108.5736G,DiCintio:2014xia}, and a Burkert profile \cite{Burkert_1995}. Both NFW-type profiles have Galactic density distributions $\rho_\chi (r)$ which take the form, as a function of galactocentric radius $r$,
\begin{equation}
    \rho_\chi (r) = \frac{\rho_0}{(r/r_s)^\gamma(1+(r/r_s)^{3-\gamma})},
\end{equation}
where we take the scale radius $r_s$ to be our local position of 8 kpc \cite{Lin_2019}, and we normalize $\rho_0$ such that $\rho_\chi (r_s) = 0.4 \text{ GeV/cm}^3$ \cite{Pato_2015}. For the standard NFW profile $\gamma = 1$, while the gNFW profile features a sharper increase in density at low galactocentric radii with a higher slope $\gamma = 1.5$.
The Burkert profile, on the other hand is cored, and takes the form 
\begin{equation}
    \rho_\chi(r) = \frac{\rho_0}{(1 + r/r_{sb})(1 + (r/r_{sb})^2)}\,,
\end{equation}
where $r_{sb}$ is the core radius. For the Burkert profile, we will take a smaller core radius than the NFW profile, such that $r_{\rm sb}=0.5$~kpc (as per Ref.~\cite{Cohen:2013ama}), to demonstrate a variety of profile choices. Note that these scale radii could of course be larger, making the dark matter density smaller towards the Galactic center.

We have assumed that the dark matter phase space distribution is not altered by our new long-range force. In the case of pure contact interactions, the gravitational potential and scattering off multiple nearby celestial bodies is not generally expected to alter the dark-matter phase space distribution. In the case of the solar system, it was shown previously that the combined effects of gravitational diffusion from the Sun, Jupiter, Mercury, and Venus, as well as elastic scattering from these bodies, leads to about the same dark matter phase-space distribution as the scenario where the celestial-body is not placed deep in the gravitational well of the solar system; see Refs.~\cite{Sivertsson:2012qj,1991ApJ...368..610G, Peter:2009mm, Peter:2009mk,Peter:2009mi}. This means that the low-velocity tail of the dark matter distribution is populated, and dark matter particles would not be sped up by the solar gravitational potential. 

However for long-range interactions, it may be possible to disrupt the dark matter phase-space distribution in the case that other celestial objects sufficiently close by alter the trajectories of virial dark matter particles. Whether this effect is relevant will depend on the particle physics parameters in question such as the coupling strength and range of the force; we neglect this possible effect and leave its consideration for future work. For celestial bodies not bound in a multiple-object system, this effect will be less relevant.

\subsection{Effective Capture Radius}
\label{accumulation}
To determine the rate of dark matter capture, we first need the effective capture radius of the body in the long-range force scenario. The capture radius is set by the maximum impact parameter $b$ such that a dark matter particle will hit the surface of the object. In the limit of purely gravitational interactions, this is the usual gravitational focusing formula,
\begin{equation}
    b^{\rm grav}_{\rm max} = R\times\sqrt{ \frac{1 +\frac{2GM}{Rv_\chi^2}}{1-\frac{2GM}{R}}}\, ,
\end{equation}
where $G$ is the gravitational constant, $M$ is the mass of the celestial body, $R$ is its radius, and $v_\chi$ is the velocity of a single dark matter particle. In the presence of a long range force between dark matter particles, this formula qualitatively changes. In this case, the maximum impact parameter can be found by starting with the radial equation of motion in the presence of a scalar potential $\Phi$ (see $e.g.$ Ref.~\cite{landau2013classical}),
\begin{align}
\label{eq: radial motion scalar mediator}
        \left(\frac{dr}{d\tau}\right)^2   &= \frac{\mathcal{E}^2}{(m_\chi+\Phi)^2} \\
        &- \left(1-\frac{2GM}{r}\right)\left(1+\frac{L^2}{r^2\,\left(m_\chi+\Phi\right)^2}\right)\,,\nonumber
\end{align}
where $\mathcal{E}$ and $L$ are the dark matter particle's energy and orbital angular momentum respectively, $r$ is the particle's radial position, and $\tau$ is proper time. The above expression assumes a Schwarzchild metric outside the celestial body. The constants of motion $\mathcal{E}$ and $L$ are fixed by the state of the dark matter particle far from the object where $\Phi(r \rightarrow \infty) = 0$,
\begin{equation}
    \begin{split}
    \mathcal{E} = m_\chi\gamma_\chi~, \\
    L = b \, m_\chi \gamma_\chi  v_\chi~,
\end{split}
\end{equation}
where $b$ is the impact parameter, and
\begin{equation}
    \gamma_\chi = \frac{1}{\sqrt{1 - v_\chi^2}}
\end{equation} 
is the relativistic gamma factor for a virialized dark matter particle. We take the scalar potential $\Phi(r)$ as
\begin{equation}
\label{eq:Phi_gauss}
    \Phi(r) = N_\chi\, V(r) = -\frac{N_\chi \alpha}{r}e^{-\lambda/r}\,,
\end{equation}
where $N_\chi$ is the number of dark matter particles accumulated by the object, and $V(r)$ is the Yukawa potential defined in Eq.~(\ref{eq:yukawa}). %

In general, an incoming dark matter particle with finite angular momentum will reach a point of closest approach.
At such a point, the radial component of the velocity vanishes, and we say the particle has hit a centrifugal barrier. For a given angular momentum, and hence a given impact parameter, the location of the centrifugal barrier is determined from the condition that Eq.~\eqref{eq: radial motion scalar mediator} must vanish. As we discuss below, when the potential has grown enough from accumulation of dark matter particles, multiple centrifugal barriers at different radii can in fact occur. This feature is not observed in the gravity only case, and would not be observed for a $1/r$ scaling in the dark potential.

\subsubsection{Case I: Low Dark Matter Abundance, Low Strength Field}

We first analyze the case of a single centrifugal barrier, when the dark potential $\Phi$ is sufficiently weak, and the force range $\lambda$ is sufficiently long, such that the $1/r$ factor in the dark potential $\Phi(r)$ limits capture. Because we are interested in a heating signal, we require that no centrifugal barrier be present at or beyond the surface of the celestial object, as otherwise this would prevent the dark matter from depositing its kinetic energy. This defines a maximum angular momentum, and therefore a maximum impact parameter $b_{\rm max}^{\rm dark}$, for which the point of closest approach matches the surface of the object  ($i.e.$ Eq.~\eqref{eq: radial motion scalar mediator} vanishes at the surface of the celestial object). Dark matter particles with angular momentum $b \leq b_{\rm max}^{\rm dark}$ thus contribute to heating, while those with $b > b_{\rm max}^{\rm dark}$ are unable to reach the object and deposit their energy. The full calculation is shown in Appendix~\ref{app: impact param}, however we predominantly work in the limit that $|\Phi| \ll m_\chi$. Under this approximation, we find that
\begin{equation}
\label{eq: bmax exp}
    b_{\rm max}^{\rm dark} \simeq R\times\sqrt{\frac{1 + \frac{2GM}{R v_\chi^2} - \frac{2\Phi}{m_\chi v_\chi^2}}{1-\frac{2GM}{R}}}~.
\end{equation}
Recall that $\Phi$ is negative so $b_{\rm max}^{\rm dark}$ grows as the accumulated dark matter number grows.

\subsubsection{Case II: Sufficiently High Dark Matter Abundance, High Strength Field}

As the number of dark matter particles inside the celestial body and therefore the dark potential grows over time, the maximum impact parameter also grows. While higher order terms in Eq.~\eqref{eq: bmax exp} need to be included in this regime, the fact that the maximum impact parameter $b_{\rm max}^{\rm dark}$ increases with increasing dark potential $\Phi$ can already be seen from Eq.~\eqref{eq: bmax exp}. As the dark potential has a Yukawa scaling, its strength is exponentially suppressed at distances larger than the force range $\lambda$. Therefore, once a large amount of dark matter is accumulated and the impact parameter grows past the force range, there is eventually a shutoff where exponential growth no longer occurs, and the growth of the dark matter number becomes sub-linear. This effect is taken into account via a second centrifugal barrier, and is only relevant in the large dark matter number regime; we derive it in Appendix \ref{app: impact param}. However, for most of our parameter space of interest, and most of the evolution time, Eq.~\eqref{eq: bmax exp} is sufficient and we focus our analytic calculations in the following sections to Case I.

\subsection{Dark Matter Energy Loss}
\label{subsec: DM energy loss}

In addition to increasing the effective capture radius, the long range force will also significantly boost in-falling dark matter particles. The boost factor as measured in the rest frame of the surface of the celestial object is given by (see Appendix \ref{app: velocity boost scalar potential})
\begin{equation}
\label{eq: gammaR}
    \gamma_R = \gamma_\chi\left[\left(1 + \frac{\Phi(R)}{m_\chi}\right) \sqrt{1-\frac{2GM}{R}}\;\right]^{-1} ~,
\end{equation}
with $\Phi$ evaluated at the surface radius $R$. 

The energy lost by dark matter in a dark matter-nucleon scattering event in the rest frame of the celestial object is then
\begin{equation}
\label{eq: DE scat}
    \Delta E_s = \dfrac{m_n m_\chi^2 \gamma_R^2 v_{R}^2}{m_n^2 + m_\chi^2 + 2\gamma_R m_\chi m_n}(1-\cos{\theta_c})\,,
\end{equation}
where $\theta_c$ is the scattering angle in the center-of-mass frame, $m_n \sim 1$ GeV is the mass of a nucleon, and 
\begin{equation}
    v_{R}^2 = 1-\frac{1}{\gamma_R^2}
\end{equation}
is the velocity of the dark matter particle at $R$ in the rest frame of the object. When $\Delta E_s$ is larger than the dark matter's initial kinetic energy, blue-shifted to the local rest frame of the celestial body,
\begin{equation}
\label{eq: KEx}
    KE_{\chi} \simeq \left(\frac{\gamma_R}{\gamma_\chi}\right)\frac{ m_\chi v_\chi^2}{2}\,,
\end{equation} 
dark matter is generally captured in one scatter. For a gravitational only potential, $\gamma_R$ is independent of $m_\chi$, resulting in $KE_\chi \propto m_\chi$ and $\gamma_R^2 v_{R}^2$ constant. This leads to multiple scatters needed in both the high and low dark matter mass regime. For non-zero coupling $\alpha$, however, $\gamma_R \appropto 1/m_\chi$, leading to both $KE_\chi$ and $m_\chi^2\gamma_R^2 v_{R}^2$ being approximately constant. Therefore, as we take $m_\chi$ smaller and smaller, we remain in the single scatter regime. On the other hand, as the dark matter mass increases, the energy loss per scatter falls off proportionally to $1/m_\chi^2$ until $\gamma_R$ begins to match up with the gravitational only value and then as $1/m_\chi$.

\subsection{Capture Probability}

To compute the capture probability of a given dark matter particle, we start by calculating the number of scatters needed for capture $n_s$, assuming an average center-of-mass scattering angle of $\theta_c = \pi/2$. Under this average angle approximation there is a simple analytic expression for $n_s$ given in Refs.~\cite{Bramante:2017xlb, Ilie:2020}, but it does not necessarily hold for highly relativistic $\gamma_R$ factors. We therefore numerically find $n_s$ by updating the dark matter kinetic energy according to Eq. (\ref{eq: DE scat}), until the condition in Eq. (\ref{eq: KEx}) is satisfied. From there we determine the probability $P_{\text{cap}}$ that the dark matter is captured,
\begin{equation}
\label {eq: Pcapture}
P_{\text{cap}} = 1 - P(n < n_s; \tau).
\end{equation}
The probability of scattering fewer than $n_s$ times is given by the sum
\begin{equation}
    P(n < n_s; \tau) = \sum_{n=0}^{n_s-1} P(n,\tau),
\end{equation} with the probability of scattering exactly $n$ times \cite{Bramante:2017xlb, Ilie:2020}, 
\begin{equation}
    \label{eq: Pn(tau)}
    P(n,\tau) = \frac{2}{\tau^2}\left(n + 1 - \frac{ \Gamma (n - 2, \tau)}{n!}\right)\,,
\end{equation}
with $\Gamma(a,b)$ the incomplete gamma function. The optical depth is
\begin{equation}
    \tau = \frac{3}{2}\frac{\sigma_{\chi n}}{\sigma_{tr}}\,,
\end{equation} 
where $\sigma_{\chi n}$ is the dark matter-nucleon scattering cross section, and
\begin{equation}
    \sigma_{tr} = \frac{\pi R^2}{N_{\rm SM}}
\end{equation}
is the transition cross section, which marks the transition from the single to multi scatter regime, and $N_{\rm SM}$ is the number of target particles inside the celestial body of interest.  Note that we have assumed that the object's density profile corresponds to a uniform sphere. When requiring a dark matter particle to pass through a smaller radius, $\frac{1}{\delta} R$ as opposed to $R$, the capture rate only decreases by the same factor $\frac{1}{\delta}$, due to the scaling with the impact parameter $b_{\rm max}$ once the dark potential dominates. As such, the majority of the brown dwarf mass residing within $\frac{1}{\delta}R$ will have only small quantitative effects on the coupling strength $\alpha$ ($\alpha \xrightarrow{\sim}\delta \alpha$) required for observable heating. We therefore expect to capture the same qualitative effects that would result from a more detailed density profile, which in any case is not a steep profile for realistic brown dwarfs.

\subsection{Accumulation Rate}
The number of dark matter particles accumulated per unit time is
\begin{equation}
\label{eq: Ndot}
    \begin{split}
        \dot{N}_\chi &= \frac{\rho_\chi}{m_\chi}\pi \langle (b_{\rm max}^{\rm dark})^2 v_\chi P_{\rm cap}\rangle \,,
    \end{split}
\end{equation}
where $\rho_\chi$ is the dark matter energy density in the position of the celestial body given by the inputs in Sec.~\ref{sec:inputs}, and $\langle \cdot \rangle$ denotes an average over the dark matter velocity distribution. The fraction of incoming dark matter particles captured depends on the single particle capture probability $P_{\text{cap}}$ given by Eq.~\eqref{eq: Pcapture}, and in the single scatter regime scales linearly with cross section $\sigma_{\chi n}$.

From Eq.~\eqref{eq: bmax exp}, note that $b_{\rm max} v_\chi$ is roughly independent of $v_\chi$. We can therefore approximate
\begin{equation}
    \label{eq: Ndot approx}
    \begin{split}
    \dot{N}_\chi \simeq \frac{\rho_\chi}{m_\chi}\pi  (b_{\rm max}^{\rm dark})^2 v_p^2 \langle \frac{P_{\text{cap}}}{v_\chi}\rangle ~,
    \end{split}
\end{equation}
 where $v_p$ is the peak of the dark matter velocity distribution (see Appendix \ref{app: Ndot} for details). We can now readily integrate Eq.~\eqref{eq: Ndot approx} by writing
\begin{equation}
    \dot{N}_\chi \simeq c_0 + c_1 N_\chi
    \label{eq:Nchi_diff}
\end{equation}
with $c_0$ and $c_1$ constants:
\begin{equation}
\label{eq: c0}
    c_0 = \kappa\left(R^2 + \frac{2GMR}{ v_\chi^2}\right)~,
\end{equation}
\begin{equation}
\label{eq: c1}
    c_1 = \kappa \left(\frac{2 R \alpha}{m_\chi v_\chi^2}\right) \, e^{-\frac{R}{\lambda}}~,
\end{equation}
\begin{equation}
    \kappa = \langle \frac{P_{\text{cap}}}{v_\chi}\rangle\left(\frac{\pi \rho_\chi v_p^2}{m_\chi}\right)\left(1- \frac{2GM}{R}\right)^{-1}~.
\end{equation}
By separation of variables we find that 
\begin{equation}
    \label{eq: Nx exp}
    N_\chi = \frac{c_0}{c_1}(e^{c_1 t}-1)~,
\end{equation}
where we have assumed the initial condition $N_\chi(0) = 0$. Note that this equation fully relies on using Eq.~\eqref{eq: bmax exp} for $b_{\rm max}^{\rm dark}$, and as such, only applies to Case I of Section \ref{accumulation}. This is an exponential accumulation phase, given the form of Eq. (\ref{eq: Nx exp}). At very late times (or very large coupling $\alpha$ relative to force length $\lambda$), eventually the exponential growth of $N_\chi$ is shut off; see Appendix \ref{app: impact param}. 

In our considered parameter space, heating occurs before an accumulated dark matter number greater than about $10^{-6} N_{\rm SM}$. Once the dark matter density becomes comparable to that of the Standard Model density, other relevant changes to the object's behavior could occur. Furthermore, once the dark matter number is high enough, the dark potential becomes comparable to the dark matter mass, and additional physical effects may become relevant that are outside the scope of this work (see Appendix \ref{app: velocity boost scalar potential}). For these reasons, we analyze only the scenario where $|\Phi| \lesssim m_\chi$. We find observable heating before this point, and note that for stronger couplings there will certainly be some observable signal, even if different to that from an increase in temperature pointed out here.

\section{Temperature from Dark Kinetic Heating}
\label{sec:kinetic}

\subsection{Instantaneous Heating}
\label{sec: FT heating}

We now calculate the instantaneous kinetic heating rate due to the long-range force. As mentioned in Sec.~\ref{sec: model set up}, we posit an additional contact interaction between dark matter and the Standard Model, and assume conservatively that an incident dark matter particle is captured solely by energy loss to scattering with Standard Model particles. The kinetic energy of a dark matter particle at the surface of a celestial body is
\begin{equation}
    KE_{\chi}(R) =  m_\chi(\gamma_R - 1),
\end{equation}
where $\gamma_R$ is given by Eq.~\eqref{eq: gammaR}. The rate of dark kinetic energy deposition is given by
\begin{equation}
    \dot{E}_\chi = KE_{\chi}(R) \times \dot{N}_\chi\,,
    \label{eq:kin}
\end{equation} 
where $\dot{N}_\chi$ is the number of dark matter particles accumulated per
unit time, as derived in the previous subsection. To relate this to the temperature $T_\chi$, we assume this energy deposition is radiated in a steady state as a blackbody,
\begin{equation}
\label{eq: SB law}
    \dot{E}_\chi= 4\pi R^2 \sigma_B T_\chi^4~,
\end{equation}
where $R$ is the radius of the object in question and $\sigma_B$ is the Stefan-Boltzmann constant. The choice of a blackbody is generally conservative, both from the point of view of telescope detection~\cite{Leane:2020wob} as well as that blackbodies cool faster than graybodies. 

Given Eqs.~\eqref{eq:kin} and~\eqref{eq: SB law}, we can determine the amount dark matter kinetic heating as a function of time 
\begin{equation}
\label{eq: T_DM}
T_{\chi} = \left[\frac{(c_0 e^{c_1t})(m_\chi(\gamma_R - 1))}{4\pi R^2 \sigma_B}\right]^{1/4},
\end{equation}
where $c_0$ and $c_1$ are given by Eq.~\eqref{eq: c0} and Eq.~\eqref{eq: c1}, and noting once again that Eq.~\eqref{eq: T_DM} applies only to Case I of Section \ref{accumulation}. Note we have assumed that the energy deposited is the dark matter's kinetic energy at the surface of the celestial body. Realistically, it will have more energy to deposit due to the fact that the bulk of the dark matter will likely sit near the center. Using Eq.~\eqref{eq: gammaR} will therefore be both conservative and more simple. The only potential issue with this assumption occurs when $\alpha \,  m_\chi \gtrsim m_\phi$, in which case the dark matter may form a compact object with large Fermi momentum, qualitatively changing the nature of the potential \cite{Wise:2014jva, Gresham:2017zqi}. However, the formation of such an object is model dependent, and only applies to relatively large dark matter mass, so we leave this possibility to future work. Another potential scenario that deviates from our assumptions occurs when a black hole forms from the captured dark matter. This happens when the dark matter that has settled at the center has grown to a critical mass for self-gravitation, which subsequently triggers collapse. The resulting black hole may in turn implode the celestial body before any substantial heating signal is achieved. This critical mass can also be model-dependent, but we note that available estimates for models closely related to our setup indicate that far more dark matter is needed than the amount that can be captured in our parameter space of interest \cite{Gresham:2018rqo}. 

Finally, Eq.~(\ref{eq: T_DM}) describes the effective temperature due to dark matter heating alone. For realistic observations one should include the temperature $T_{\rm SM}$ from Standard Model backgrounds as well. Energy deposition from separate sources can be added linearly, meaning the total temperature is calculated as
\begin{equation}
\label{eq: T_obj}
    T_{\text{obj}} = \left[{T_{\chi}^4 + T_{\rm SM}^4}\right]^{1/4}~,
\end{equation}
with $T_{\rm SM}$ the temperature expectation from Standard Model sources in the absence of dark matter heating. We have assumed in Eq. \eqref{eq: T_obj} that all of the DM kinetic energy goes into an increase in the SM temperature of the object. While some small amount of the energy can go into increasing the radius of the object at the edge of our parameter space, this is at most a correction that deviates from the expectation under constant radius by only a factor of $\lesssim\sqrt{1.5}$~\cite{Macintosh:2023pc,Croon:2024waz}. For the bulk of our parameter space, it is even smaller and negligible. Assuming a constant radius is therefore a good approximation for our parameters of interest.

\subsection{Thermal Equilibrium Timescales}
\label{sec:therm}

It is important to note that there is a potentially significant time delay between dark matter energy deposition and the stage where the celestial body itself reaches thermal equilibrium, which has not always been calculated in previous dark matter heating works. This effect is distinct from dark matter thermalization timescales, or dark matter capture and annihilation equilibrium. Physically it arises because the Standard Model particles take time to reach the new temperature, due to the heat capacity of the material. This means that objects need to be sufficiently old before effects of dark matter energy injection are relevant, and very young objects will fail this condition. We will show however for our dark kinetic heating parameter space of interest, thermal equilibrium of the celestial body is achieved in a sufficiently short timescale, as we are not primarily concerned with the early stages of the celestial body.

We estimate the time it takes for a celestial object to be kinetically heated to a final steady-state temperature starting from the equation,
\begin{equation}
    C_v \frac{dT}{dt} = \dot{E}_{\chi}(t) - \dot{E}_{\rm cool}(t)\,,
    \label{eq:therm_evol_main}
\end{equation}
where $C_v$ is the constant-volume heat capacity of the celestial object. The dark matter heating rate energy deposition rate $\dot{E}_{\chi}$ is given by Eq.~\eqref{eq:kin}. We will discuss our assumptions for heat capacity and the resulting thermal equilibration timescales for a few celestial objects in the upcoming sections.

\section{Exoplanet and Brown Dwarf Kinetic Heating}
\label{sec:bd}

Equipped with a framework for dark kinetic heating in the presence of long-range forces, we now apply our calculation to an example class of low-escape velocity objects: exoplanets and brown dwarfs. As elucidated in the introduction, these objects are advantageous for detection due to their large surface areas, leading to larger luminosities for fixed temperatures, but they also have an extensive upcoming search program with $e.g.$ JWST, Roman, and Rubin Telescopes. Tests far into the Galactic center also may allow potential probes into the largely unknown dark matter distribution in our Milky Way Galaxy. However, they have previously only been studied in the context of dark matter annihilation heating~\cite{Leane:2020wob,Benito:2024yki}. We will show now that the discovery potential for this search is greatly enhanced when considering broader classes of particle physics models that produce the dark kinetic heating effect. We will consider other low-escape velocity objects in Sec. \ref{sec: other objects} that, depending on the model parameters, may also be of complementary interest.

\begin{figure}[t]
    \centering
%    \hspace{-5mm}
\includegraphics[width=\linewidth]{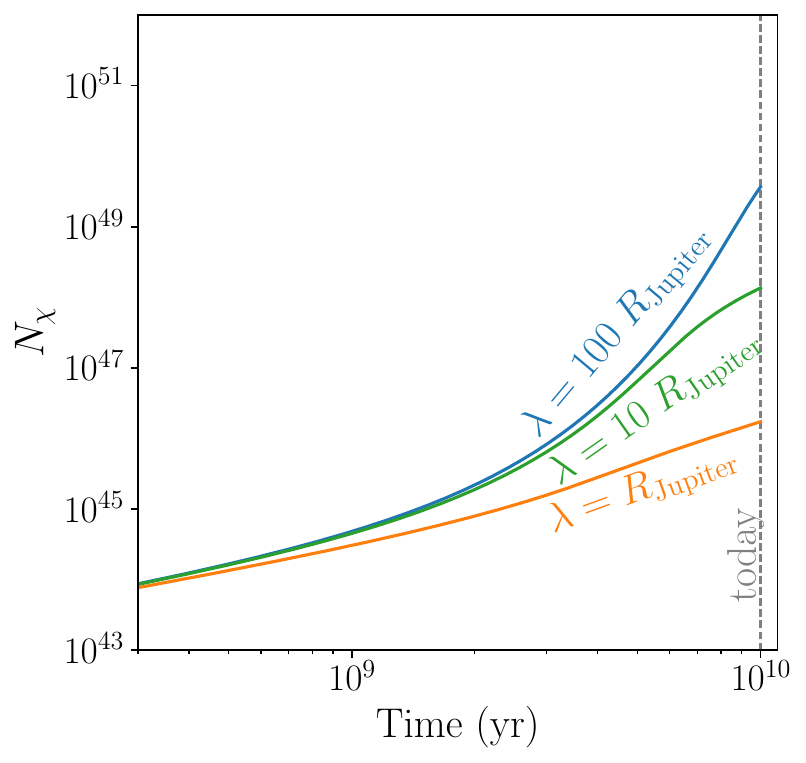}
    \caption{Number of accumulated dark matter particles with GeV mass within a 10 Gyr old, 55 Jupiter mass brown dwarf with Jupiter's radius, at local position over time. The blue ($\lambda =  100 \;R_{\rm Jupiter}$), green ($\lambda = 10 \;R_{\rm Jupiter}$), and orange ($\lambda = R_{\rm Jupiter}$) lines show three different example values of the force range $\lambda$ as multiples of the brown dwarf's radius, all with the same coupling strength $\alpha = 3\times 10^{-27}$. 
    } 
    \label{fig: Nx_over_time_BD}
\end{figure}

\subsection{Sensitivity to Coupling Strength and Force Range}
We begin by investigating the effective dark matter heating of a brown dwarf as a function of both interaction strength and force range. We will start by assuming that $99\%$ of dark matter particles that hit the surface are captured, in order to probe approximately maximum sensitivity to coupling strength $\alpha$ and force range $\lambda$, before analyzing sensitivity to arbitrary captured fractions and therefore the scattering cross section $\sigma_{\chi n}$. We now demonstrate the physical behavior of dark matter accumulation over time, for some somewhat arbitrary benchmark parameters. 

Figure~\ref{fig: Nx_over_time_BD} shows the number of dark matter particles $N_\chi$ as a function of time $t$, within an example brown dwarf at local position with the same radius as Jupiter, and 55 times the mass, for a few values of the force range $\lambda$ and an example fixed coupling strength $\alpha = 3\times 10^{-27}$. Here we have taken $m_\chi = 1$ GeV for simplicity. Accumulation starts out quite slow, matching the typical gravitational capture, then begins to increase exponentially as the dark matter number and potential grow, before eventually slowing down when the effective capture radius reaches becomes comparable to the size of the force range. As can be expected, the change in brown dwarf temperature will not be significant until the brown dwarf's dark matter number has reached the exponential phase of accumulation and the potential has grown appreciably. Therefore, the largest signals clearly arise for force ranges larger than the radius of the object. We see that we are sensitive to extremely small couplings of the long-range self-interaction, though this is aided somewhat by the fact we have also considered an additional contact interaction for energy loss. Without the contact interaction, the couplings probed here would indeed be much larger, but are highly model dependent. 

\begin{figure}[t]
    \centering
\includegraphics[width=\linewidth]{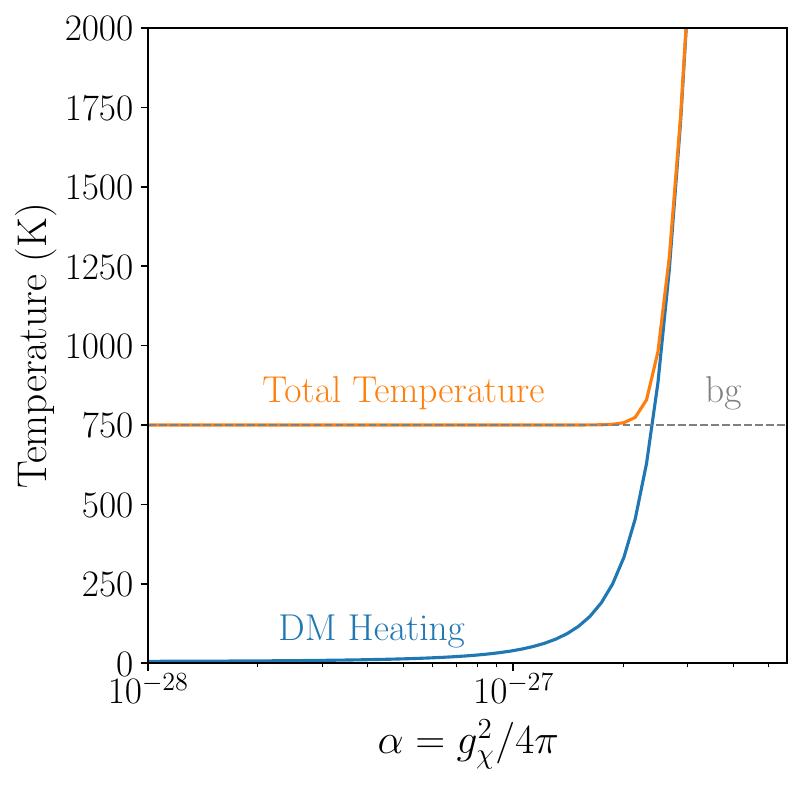}
    \caption{Dark kinetic heating of a 10 Gyr old, 55 Jupiter mass brown dwarf with Jupiter's radius, at local position as a function of coupling strength $\alpha$ with fixed force range $\lambda$ equal to the 100 times brown dwarf's radius. The gray dashed line ``bg" indicates a background temperature of 750K, the blue line shows the dark kinetic heating, and the orange line displays the total brown dwarf temperature.}
    \label{fig: T_over_alpha_BD}
\end{figure}

Figure~\ref{fig: T_over_alpha_BD} shows the dark kinetic heating today of our benchmark brown dwarf as a function of $\alpha$, along with the Standard Model background ``bg" ($T_{SM}= 750$ K) and the total combined temperature. 
In order to decouple the analysis of varying $\alpha$ from varying $\lambda$, we have shown an example value of $\lambda$ equal to 100 times the brown dwarf's radius.
We notice that the dark kinetic heating follows a similar trend to the dark matter number $N_\chi$ over time, starting out slow and then hitting an exponentially growing period. For smaller force ranges, we would also have seen the exponential shut off, but for the plot range shown, we never reach this phase for the $\lambda$ chosen, as $\Phi \lesssim m_\chi$. The left side of this plot begins with a slightly non-zero heating value, which comes from gravitational dark matter capture even in the limit of negligible dark potential -- this is dark kinetic heating without long-range forces, and shows why dark kinetic heating previously was considered irrelevant for low-escape velocity objects such as brown dwarfs, as it is so far below the background temperature.

\begin{figure}[t]
    \centering
\includegraphics[width=\linewidth]{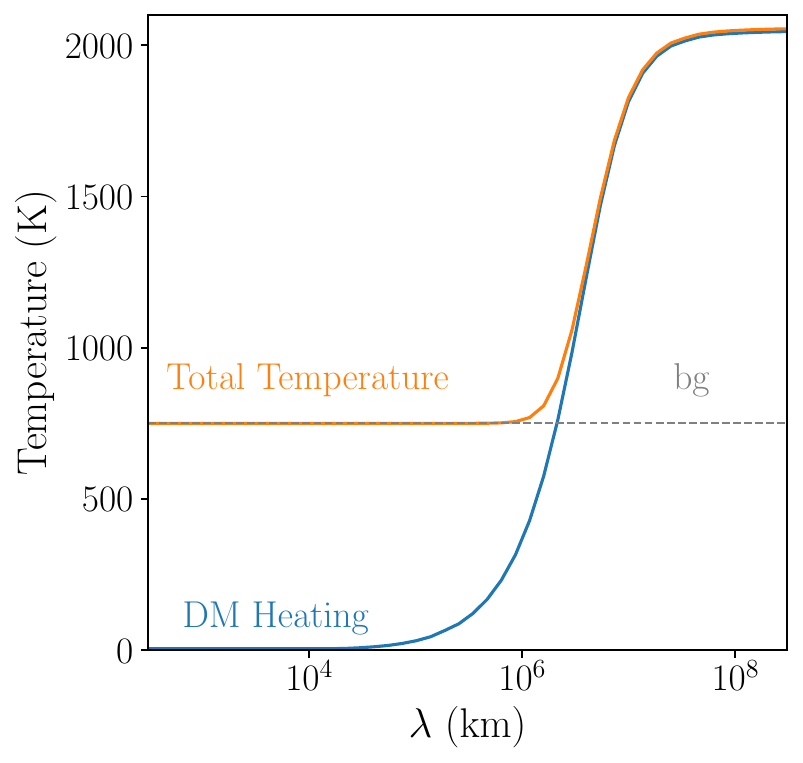}
    \caption{Dark kinetic heating of a 10 Gyr old, 55 Jupiter mass brown dwarf with Jupiter's radius, at local position as a function of the force range $\lambda$, with fixed coupling strength $\alpha = 2.85 \times 10^{-27}$.  The gray dashed line ``bg" indicates a background temperature of 750K, the blue line shows the dark kinetic heating, and the orange line displays the total brown dwarf temperature.}
    \label{fig: T_over_lambda_BD}
\end{figure}

Figure~\ref{fig: T_over_lambda_BD} demonstrates dark kinetic heating of a brown dwarf as a function of force range $\lambda$. Here we have again taken $m_\chi = 1$ GeV for simplicity. We choose a fixed value of the coupling $\alpha = 2.85\times 10^{-27}$ as we vary the force range. This value is chosen as the maximum value of $\alpha$ such that heating shown is in a regime where energy deposition dominantly causes a temperature increase, rather than radius increase, of the object. We emphasize that this coupling $\alpha$ is chosen for demonstrative purposes, and any $\alpha \gtrsim 2.85 \times 10^{-27}$ would have strong heating signals as well. This plot shares features similar to Fig. \ref{fig: T_over_alpha_BD}, where the left side asymptotes to the same value in the limit of negligible dark potential, and then an exponential increase in heating occurs as we increase the force range. Unlike the case of increasing the coupling $\alpha$, however, heating eventually saturates to a set value as we increase the force range large enough. This is because after a certain range ($\lambda \sim \mathcal{O}(10^{7})$ km for the example parameters we have chosen), the long range force is already behaving as essentially infinite relative to other parameters of the theory. 

\begin{figure*}[t]
\centering
    \includegraphics[width=0.47\linewidth]{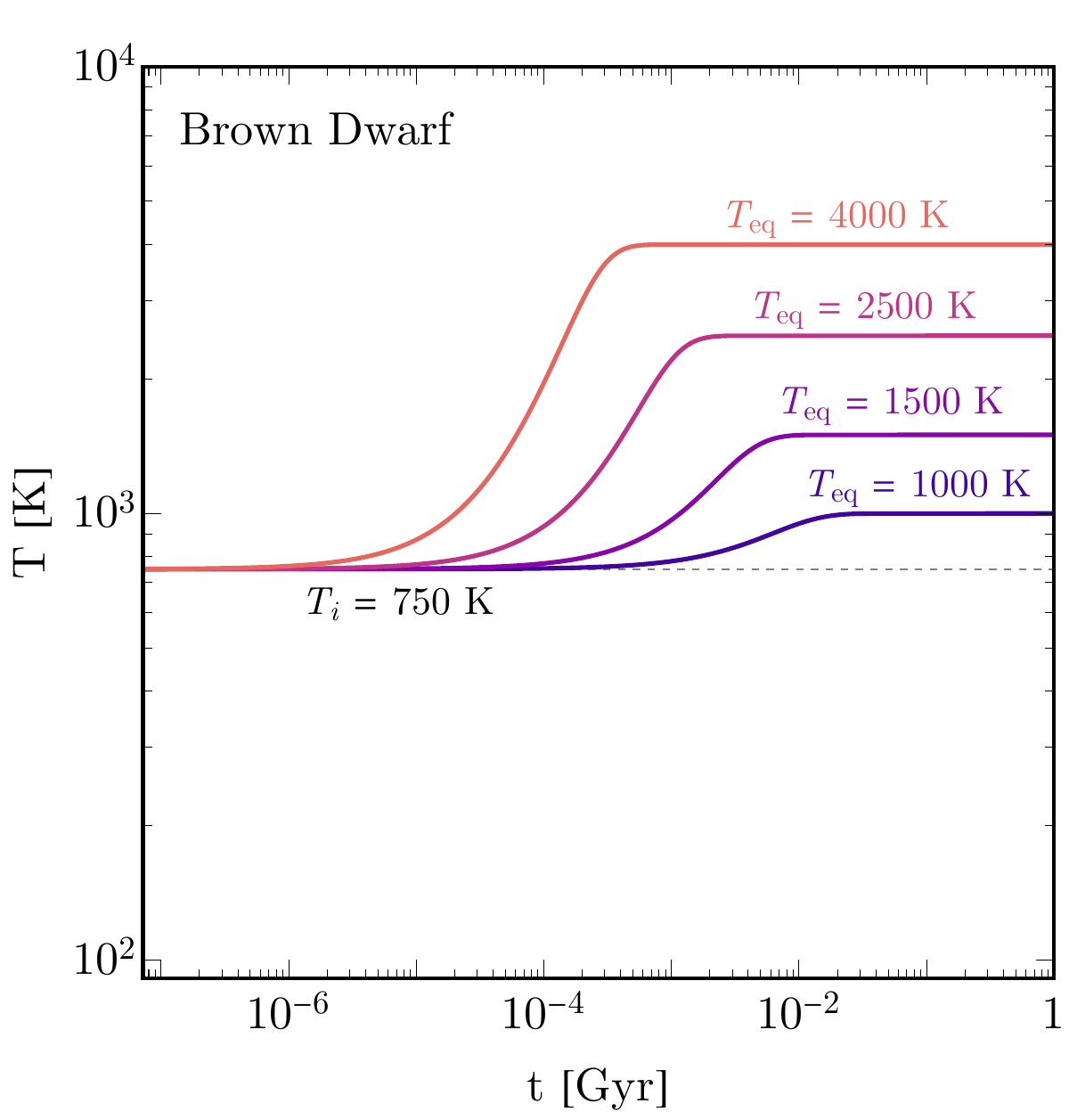}\hspace{8mm}
   \includegraphics[width=0.47\linewidth]{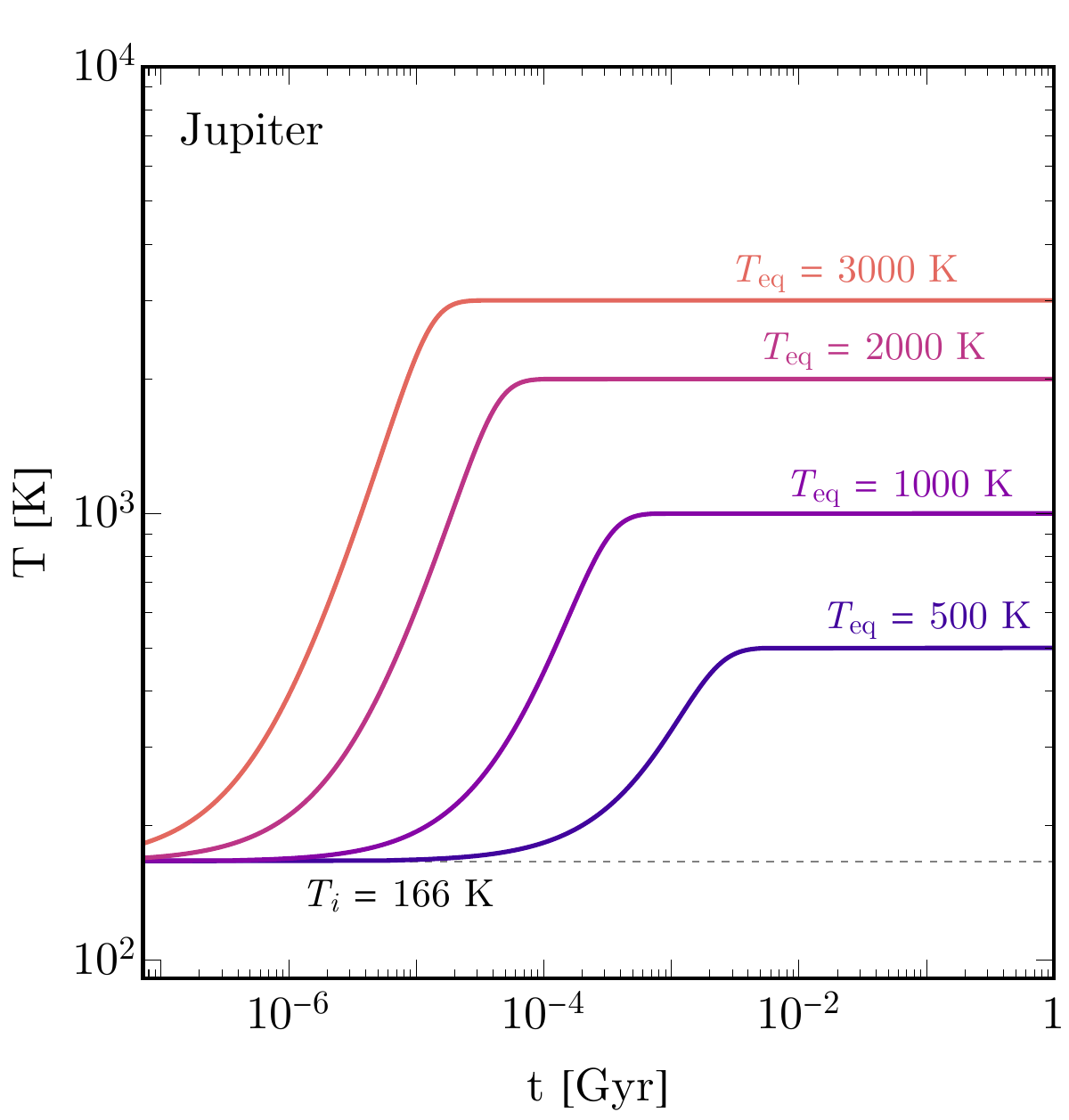}
\caption{Temperature evolution for a benchmark 55 Jupiter-mass brown dwarf and Jupiter, assuming fixed energy injection rates. The dashed line indicates the initial temperature $T_i$ assumed. Each colored label shows the final equilibrium temperature $T_{\rm eq}$ reached for the given line.} 
\label{fig:BD_heat_timescale}
\end{figure*}

An interesting feature of Fig.~\ref{fig: T_over_lambda_BD} is the value of the force range for which the exponential increase in heating occurs. The exponential region is centered roughly around $\sim O(10^6)$ km, which is about two orders of magnitude above the brown dwarf radius. When the force range is not much more than the brown dwarf radius, some additional heating begins compared to the purely gravitational picture, but it is still not much, since the force is exponentially suppressed everywhere outside of the object. When the force range becomes much larger than the brown dwarf's radius however, the long range force starts having a strong effect on heating as surrounding dark matter particles feel an unsuppressed dark potential. For force ranges significantly larger than the brown dwarf's radius, larger objects might offer more sensitivity given the larger mass and radius for starting capture, so long as the force range is still larger than the celestial-body radius. We might therefore expect that the optimal astrophysical target for a given force range is one for which the force range is larger than the radius of that object, but not significantly larger than the radius of any other. We investigate this idea further in the context of other objects than brown dwarfs, in the upcoming Section~\ref{sec: other objects}. 

\subsection{Brown Dwarf and Exoplanet Thermal Equilibrium}

For the benchmark dark kinetic heating rates established in the previous section, we now show that these objects do reach thermal equilibrium faster than the rate at which the dark kinetic energy deposition changes appreciably, such that the evolution of the temperature is effectively instantaneous.

For exoplanets and brown dwarfs, we assume for simplicity an ideal gas heat capacity of the form
\begin{equation}
    C_v = \frac{N_{\rm SM}}{\Gamma - 1}~,
    \label{eq:Cv_main}
\end{equation}
where $\Gamma$ is the adiabatic index, which for most gases ranges between $9/7$ to $5/3$. In our estimates, we use $\Gamma = 7/5$ corresponding to an ideal diatomic gas. Note that, depending on the temperature, the hydrogen molecules in brown dwarfs and gas giants may be dissociated into a monoatomic gas. However, we use the diatomic heat capacity in our computations. This is conservative since diatomic gases have a larger heat capacity compared to monoatomic gases, resulting in a slightly larger estimate of the thermal equilibrium timescale. Depending on the temperature and pressure within the object, other exotic phases with different heat capacities may arise, but their presence is model-dependent and so we do not consider these for simplicity. We also assume these objects cool down as blackbody radiators, $i.e.$ following Eq.~\eqref{eq: SB law}.

Figure~\ref{fig:BD_heat_timescale} shows the temperature evolution of a brown dwarf with mass $M_{\rm BD} = 55 M_{\rm Jupiter}$ and radius $R = R_{\rm Jupiter}$, and a benchmark Jupiter (applicable also to exo-Jupiters), for various fixed energy injection rates. These were chosen to approximately span the dark kinetic heating rates obtained in our parameter space of interest, though they could be mapped to any model scenario ($e.g.$ dark matter annihilation) that has the same goal equilibrium temperature as labeled. In all cases, one can see that brown dwarfs and Jupiters should reach the equilibrium temperature on a timescale short compared to the dark kinetic heating build-up.

\begin{figure*}[t]
    \centering
\includegraphics[width=\textwidth]{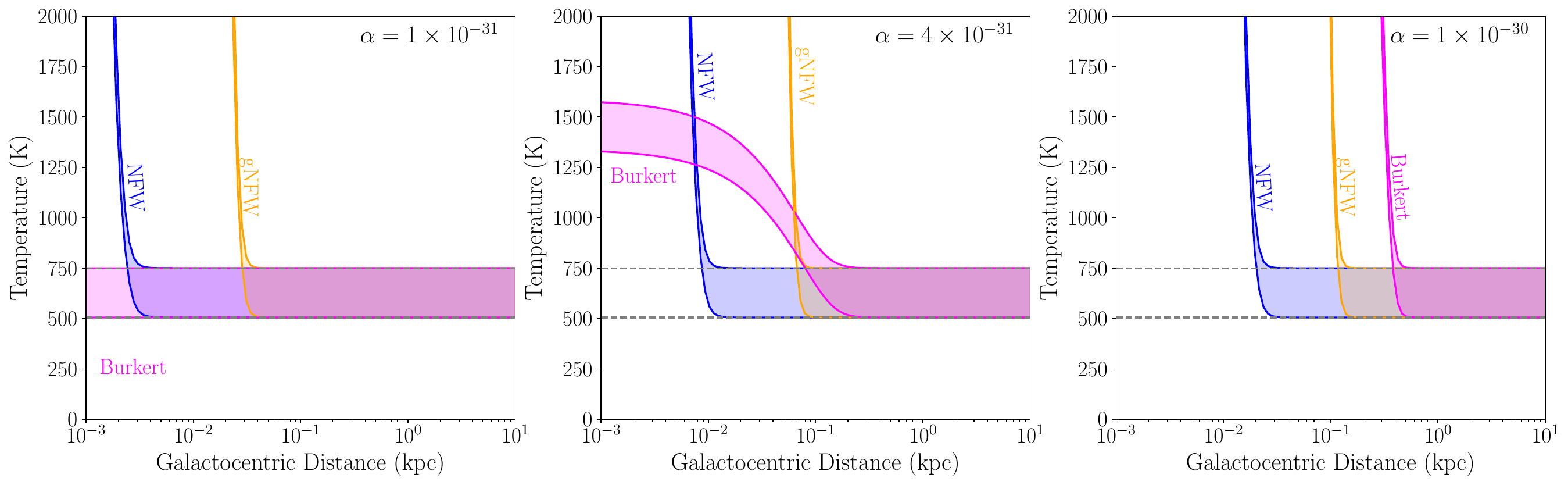}
\caption{Dark kinetic heating of a 10 Gyr old brown dwarf with the radius of Jupiter. The lower line for each color represents a 35 Jupiter mass brown dwarf, and the upper line represents a 55 Jupiter mass brown dwarf; the shaded region between them corresponds to intermediate mass values. The lower dashed line represents the 500 K Standard Model background temperature of a 35 Jupiter mass brown dwarf, and the upper dashed line represents the 750 K Standard Model background temperature of a 55 Jupiter mass brown dwarf. Three example values of the interaction strength $\alpha$ are shown, all with an example force range of 100 times the size of the body's radius.}
    \label{fig: galacto-radial_heating_BD}
\end{figure*}

\subsection{Observational Expectations}
\label{sec: observations}

 We now examine the expected temperatures of brown dwarfs (and brown dwarf-like exoplanets such as Super-Jupiters) as a function of distance from the center of the Galaxy, under the presence of dark kinetic heating.  

Figure~\ref{fig: galacto-radial_heating_BD} shows temperature plots for two different mass brown dwarfs with 35 and 55 Jupiter masses and Jupiter's radius, as a function of galactocentric distance, for example values of the interaction strength and force range. For the 35 Jupiter mass brown dwarf we use a Standard Model background temperature of 500 K, and for the 55 Jupiter mass brown dwarf we use a Standard Model background temperature of 750 K~\cite{Leane:2020wob}. We see that brown dwarf temperature as a function of galactocentric distance is predicted by specific Galactic dark matter density profiles, and therefore measuring such temperatures could provide insight to the unknown dark matter density, as was originally pointed out in the context of the dark matter annihilation only scenario in Ref.~\cite{Leane:2020wob}. In the case of dark kinetic heating however, we see that the rates can be substantially larger. In fact, for different interaction and force range values, dark kinetic heating can even be appreciable in the local position, $e.g.$ if larger couplings were chosen than those shown in this figure. Indeed, similar plots to those shown in Fig.~\ref{fig: galacto-radial_heating_BD} can be made for different values of the force range and coupling strength, allowing us to probe the full model with the observation of more distant objects, as well as local temperature measurements.

\subsection{Constraints from WISE and JWST Data on a Super-Jupiter Exoplanet}

Given the large dark kinetic heating signals possible at very small coupling parameters, it is pertinent to consider how dark kinetic heating can already lead to exclusions of the dark matter parameter space with current observations. While there are a range of new and upcoming infrared and optical telescopes such as JWST, Roman, and Rubin as discussed earlier, there are already existing measurements with sensitivity to our scenario. 

The coldest observed large exoplanet is the Super-Jupiter WISE 0855-0714. A Super-Jupiter is effectively an object like Jupiter, but more massive. As they are more massive, they can be more dense, leading to a superior probe of the dark matter scattering cross section compared to Jupiter in our own solar system. The Super-Jupiter WISE 0855-0714 was observed by the Wide-field Infrared Survey Explorer (WISE), which launched in 2009 and has been observing the Universe in infrared ever since. It was also recently observed by JWST with its Near-Infrared Spectrograph (NIRSpec), which launched in 2021. We use the temperature measurement and uncertainty from Ref.~\cite{Kirkpatrick_2021} of $250 \pm 50$ K, which is consistent with JWST's spectral measurement matching atmospheric models corresponding to a 285 K object~\cite{luhman2023jwst}. WISE 0855-0714 has a mass between 1 and 10 times that of Jupiter, a radius of $0.9-1\;R_{\rm Jupiter}$, an age of $0.3-6$ Gyr, and a distance approximately 2.28 pc from Earth~\cite{Luhman_2014, Leggett_2017, Kirkpatrick_2021, luhman2023jwst}. Its low temperature is already in conflict with some dark matter model parameters which lead to dark kinetic heating.

\begin{figure}[t]
    \centering
\includegraphics[width=\linewidth]{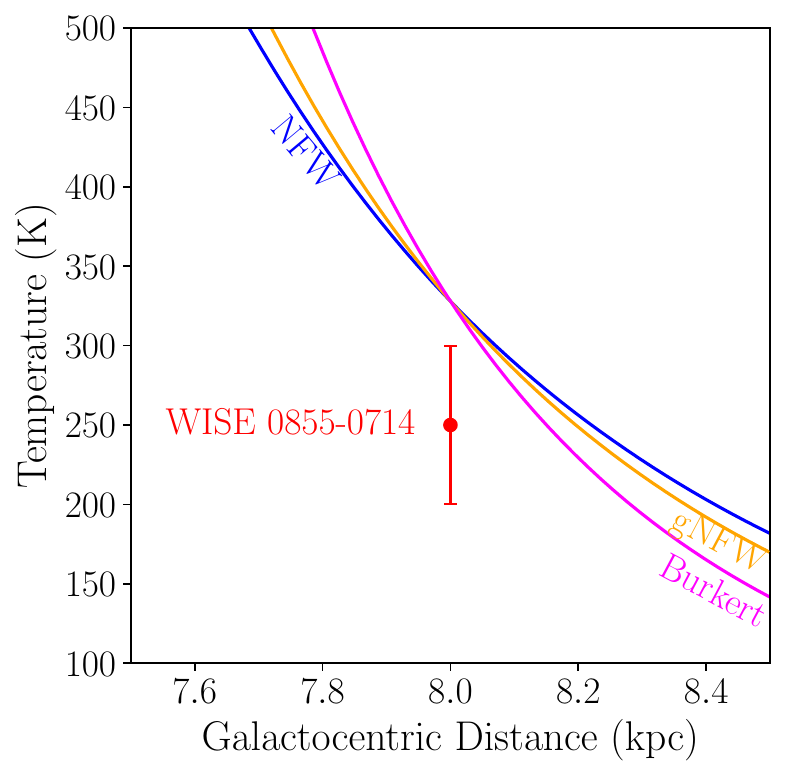}
    \caption{Dark kinetic heating of a 0.3--6 Gyr old, 1--10 Jupiter mass Jovian exoplanet with 0.9--1 Jupiter radius. We assume all heating is due to dark kinetic heating and plot the temperature of WISE 0855-0714~\cite{Kirkpatrick_2021} in red. See text for applicable dark kinetic heating parameters.}
    \label{fig: WISE}
\end{figure}

Figure~\ref{fig: WISE} shows the temperature of WISE 0885-0714 and its associated uncertainties, contrasted with expected dark kinetic heating rates for such an object as a function of galactocentric distance, conservatively neglecting the Standard Model contribution to the temperature. As the dark kinetic heating rate for example parameters shown is in excess of the measured temperature, dark matter models with interaction strengths of $\alpha \geq 8.6\times 10^{-26}$ and force range lengths of $\lambda \geq 90 \,R_{\rm Jupiter}$ are already ruled out by the existence inferred temperature of WISE 0885-0714, if the low end estimates of $M_{\rm Jupiter}$ mass, 0.9 $R_{\rm Jupiter}$ radius, and 0.3 Gyr age are taken. For the highest end WISE 0885-0714 estimates of 10 $M_{\rm Jupiter}$ mass, $R_{\rm Jupiter}$ radius, and 6 Gyr age, this corresponds to an even stronger constraint at interaction strengths of $\alpha \geq 3.7\times 10^{-27}$ and force range lengths of $\lambda \geq\, 100\,R_{\rm Jupiter}$ being ruled out. The coupling $\alpha$ and force range $\lambda$ can be raised and lowered inversely to get a similar bound. These constraints on the force range and coupling are for cross sections corresponding to the maximum capture rate in the Super-Jupiter, which are larger than about $10^{-35}~$cm$^2$ at 1 GeV dark matter mass.

Intriguingly, for dark matter parameters that would give the exact amount of heating required to sustain the 250 K temperature, one could also question whether the Super-Jupiter WISE 0885-0714 is actually significantly older and represents a positive dark matter signal, for alternate dark matter parameters. Indeed, it is important to note that the mass, temperature, and age of this Super-Jupiter are not measured independently, but rather inferred based on spectral data \cite{Leggett_2017}. This inference includes Standard Model-only assumptions, and therefore would change under the addition of dark kinetic heating. In any case, limits can however be drawn by taking conservative choices for any inferred Standard Model properties. A true positive signal of dark kinetic heating may look like a statistically significant number of objects with temperature and luminosity measurements that do not match standard cooling theory, nor standard expectations for any other object of the same age. This has been established in another work that focused on DM-heating of main-sequence stars~\cite{John:2024thz}, where a new population is expected to appear that is observationally inconsistent with the SM. To further distinguish a brown dwarf heated by dark matter from similar objects, such as a younger red dwarf, one can also use lithium concentrations to independently determine the age of the object \cite{Basri1998}. In addition, microlensing measurements can provide an independent measure of brown dwarf masses in some cases~\cite{han2017ogle}. Finally, observation of objects which increase with temperature towards the galactic center, in a way that mirrors an expected DM distribution, would be a smoking gun signal of dark matter heating.

\begin{figure}[t]
    \centering
\includegraphics[width=\linewidth]{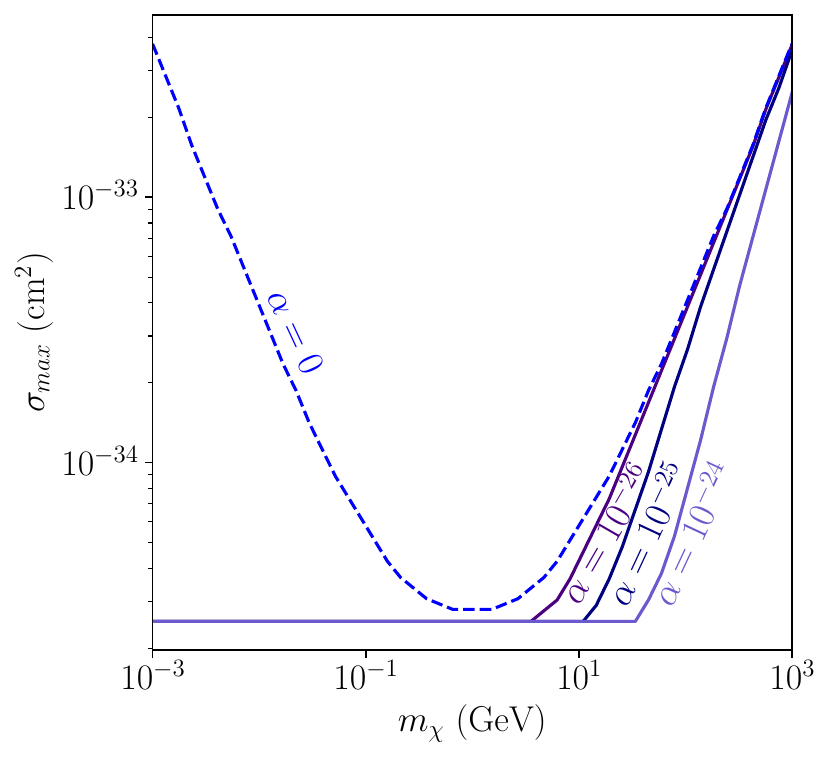}
    \caption{Cross section required for 99\% of dark matter particles that pass through a 55 Jupiter mass brown dwarf with the radius of Jupiter to be captured. An example fixed dark matter number of $N_\chi = 5\times10^{-12}N_{\rm SM}$ is used here, where $N_{\rm SM}$ is the number of Standard Model nucleons in the object. A fixed force range $\lambda$ equal to 100 times the brown dwarf's radius is also chosen. The dashed blue line represents dark interaction strengths $\alpha = 0$, displaying the required cross section in the absence of a long range force, and the solid lines represent different values of $\alpha$. The values of $N_\chi$ and $\alpha$ are arbitrarily chosen to demonstrate the effect on $\sigma_{\rm max}$ of a long range force, as $N_\chi$ is a function of time for any chosen value of $\alpha$.}
    \label{fig: sigma max}
\end{figure}

\begin{figure}[t]
    \centering
\includegraphics[width=\linewidth]{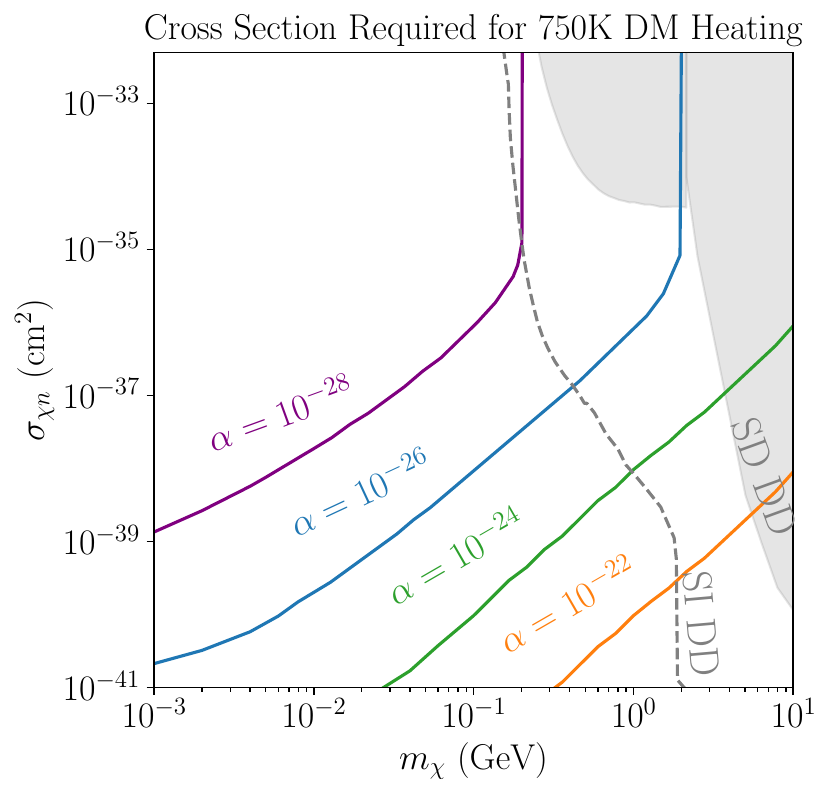}
    \caption{Cross section required for dark matter to kinetically heat a 10 Gyr old, 55 Jupiter mass brown dwarf with Jupiter's radius to 750 K. Various values of interaction strength $\alpha$ are plotted, all with force ranges $\lambda$ set to 100 times the radius of the brown dwarf. Complementary constraints from direct detection on spin-independent scattering ``SI DD" and spin-dependent scattering ``SD DD" are also shown, see text for details.}
    \label{fig: sigma sensitivity}
\end{figure}

\subsection{Scattering Cross Section Sensitivity}
\label{sec: cross section}

Figure~\ref{fig: sigma max} shows the cross section corresponding to the assumption that any dark matter particle passing through a benchmark brown dwarf is captured with 99\% probability, $\sigma_{\rm max}$, which we had so far assumed. After some dark matter has been accumulated, the required cross section for 99\% capture changes, and Fig.~\ref{fig: sigma max} also shows $\sigma_{\rm max}$ as a function of dark matter mass for a few values of the interaction strength $\alpha$ at a fixed dark matter accumulation number. The transition cross section of this object, corresponding to the switch between the single and multi scatter regimes, is at about $2\times10^{-35}$~cm$^2$. In the case of a gravitational potential only, $i.e.$ $\alpha=0$, we observe the usual expectation for a local brown dwarf, which is that it usually requires multiple scatters to capture all of the incoming dark matter. The fact it requires multiple scatters is clear because the $\sigma_{\rm max}$ value is always above the transition cross section of about $2\times10^{-35}$~cm$^2$, and approaches it only when the dark matter mass becomes comparable to the target mass of around 1 GeV, when scattering is kinematically efficient. 

Interestingly, we see that as the interaction strength of the long-range force increases, the value of the cross section which captures the bulk of the dark matter rapidly approaches the transition cross section, $i.e.$ we see the cross sections going flat with dark matter mass. This means that the effect of the long-range force is to greatly increase the capture efficiency, and therefore greatly increase the regime for which $\sigma_{\rm max}$ is the cross section required for the dark matter to scatter once, particularly in the low dark matter mass regime. As dark matter mass increases above 1 GeV the long range force becomes weaker until $\sigma_{\rm max}$ matches the gravity only case. This effect occurs because for heavier dark matter the same potential difference will correspond to slower incoming speeds, as per the discussion in Section \ref{subsec: DM energy loss}. Fig.~\ref{fig: sigma max} was generated assuming a fixed dark matter number of $N_{\chi} = 5\times 10^{-12} N_{\rm SM}$, chosen simply to display the effect of the long range force on $\sigma_{\rm max}$ at some arbitrary snapshot in time. Note that the interaction values $\alpha$ on this plot are not in direct parallel to those shown in previous Figs.~\ref{fig:  Nx_over_time_BD}--\ref{fig: T_over_lambda_BD}, as this plot is showing a particular snapshot in time for the dark matter number. 

Figure~\ref{fig: sigma sensitivity} shows the minimum cross section to obtain 750 K or higher of dark kinetic heating as a function of dark matter mass for varied interaction strengths, for a benchmark brown dwarf. The value of 750 K is chosen as an example, as it is sufficient to overcome the background Standard Model temperature of this benchmark brown dwarf, and be detectable by telescopes. We also show complementary bounds on the dark matter scattering cross section from direct detection experiments. These include the strongest limits arising from spin-dependent proton-dark matter scattering, which are from CRESST-III~\cite{CRESST:2022dtl} and PICO-60~\cite{PICO:2019vsc}. We also show the strongest spin-independent nucleon-dark matter scattering bounds, from CRESST~\cite{CRESST:2019jnq,CRESST:2022lqw}, DarkSide~\cite{DarkSide:2018bpj}, XENON-nT~\cite{XENON:2023cxc}, and LZ~\cite{LZ:2022ufs}. These constraints may be stronger or weaker in the presence of a long range force that qualitatively alters the escape velocity at Earth's surface ~\cite{Davoudiasl:2017pwe,Davoudiasl:2020ypv}. However, for the parameters shown in Fig. \ref{fig: sigma sensitivity}, Earth does not accumulate a sufficient number of dark matter particles for the direct detection bounds to change. There are other potential complementary astrophysical bounds in this parameter space, however they are not directly comparable to our scenario as they are set by different particle model assumptions, such as the inclusion of dark matter annihilation. For completeness examples of these bounds are from altered S-star evolution due to dark matter~\cite{John:2023knt}, dark matter induced ionization of Jupiter's ionosphere~\cite{Blanco:2023qgi}, or gamma-ray measurements of celestial objects or populations~\cite{Leane:2017vag,HAWC:2018szf,Nisa:2019mpb,Leane:2021tjj,Leane:2021ihh,Acevedo:2023xnu,Linden:2024uph}.

Overall, in Fig.~\ref{fig: sigma sensitivity} we see that dark kinetic heating can lead to a substantial increase in the dark matter parameter space that can be probed, exceeding cross sections even down to $10^{-41}$\,cm$^2$. In the figure, we see that increased interaction strength or decreased dark matter mass leads to superior cross section sensitivity. This is expected because the dark potential scales with interaction strength and dark matter number, but not dark matter mass itself (see Eq.~\eqref{eq:Phi_gauss}). Above a certain dark matter mass, sharp cut offs are observed. This occurs because sensitivity disappears when capturing all of the incoming dark matter is still not sufficient to achieve the fixed heating of 750 K ($i.e.$ the maximum capture cross section is reached). On the other hand, as the mass decreases, sensitivity stops increasing as rapidly due to the required initial number of scatters for capture going up. Once an appreciable number of dark matter particles have been accumulated, most particles can be captured in a single scatter as is shown in Fig.~\ref{fig: sigma max}. However in the early stages of capture when there are not many dark matter particles and the dark potential is weak, capture of dark matter with mass away from the target mass will occur dominantly from a smaller portion of the low-velocity tail of the Galactic dark matter distribution. While we show only the cross section required to achieve 750 K of dark kinetic heating as an example in Fig.~\ref{fig: sigma sensitivity}, very small increases in the scattering cross section lead to significant increases in heating due to the exponential dependence on the capture rate, consistent with Fig. \ref{fig: T_over_alpha_BD}.

Lastly, note in Figs.~\ref{fig: sigma max} and \ref{fig: sigma sensitivity} that we have ignored any potential evaporation effects. Evaporation occurs when the dark matter is up-scattered by the celestial body matter, such that it can overcome the object's escape velocity and leave the object, removing any potential dark matter signal. While long-range self-interactions will mitigate evaporation once some dark matter has been accumulated, the rapid evaporation of very light dark matter might negate the possibility of any build-up at all. However, there are a wide range of scenarios where evaporation does not generally occur, even at very light dark matter masses. For example, when the field $\phi$ also couples to nucleons evaporation is greatly mitigated \cite{Acevedo:2023owd}, and can be suppressed by multiple orders of magnitude below usual contact interaction expectations. We expand on this point in Appendix~\ref{app:dm_evap}. We therefore simply note that some regions of Fig.~\ref{fig: sigma sensitivity} might require more specific ingredients than the minimal toy model adopted here.

\section{Other Astrophysical Targets}
\label{sec: other objects}

We now investigate the complementarity of dark kinetic heating of multiple astrophysical objects, all of which have not previously been considered for detectable dark kinetic heating due to their low escape velocities. We will demonstrate that depending on the parameter space of interest, certain objects offer better discovery potential or constraints than others. We will also show that the benchmark kinetic heating signals calculated in the previous sections for brown dwarfs are not ruled out already by other objects. We will consider three targets in addition to brown dwarfs/exoplanets: white dwarfs, nuclear-burning stars such as our Sun, and the Earth. We also comment on this effect in neutron stars.

\begin{figure*}
    \centering
\includegraphics[width=0.495\linewidth]{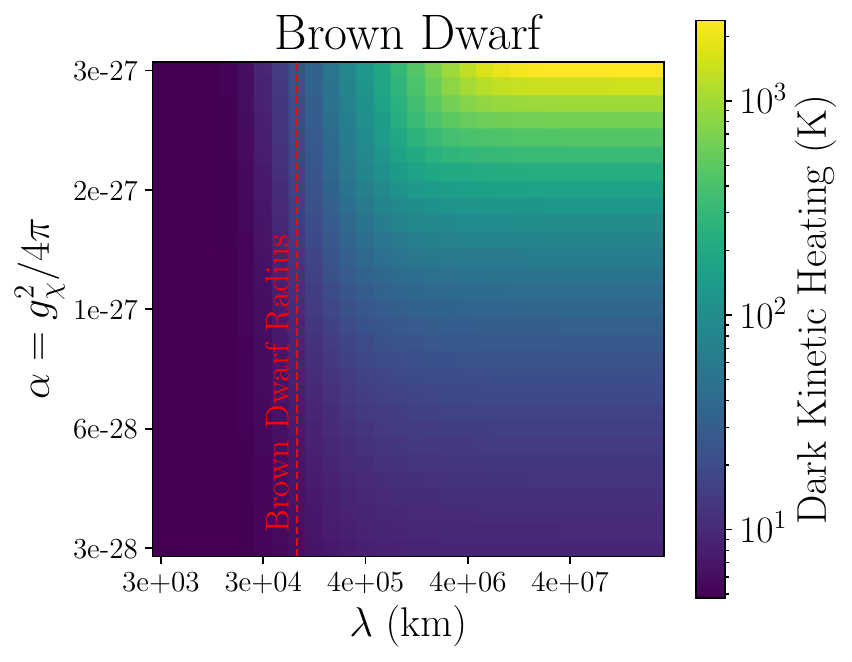}
\includegraphics[width=0.495\linewidth]{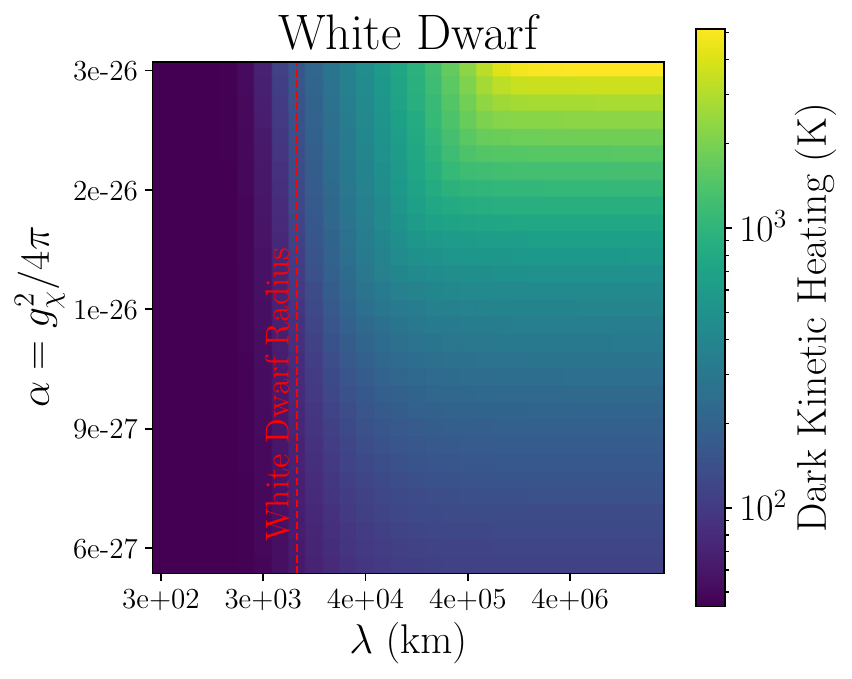}
\includegraphics[width=0.495\linewidth]{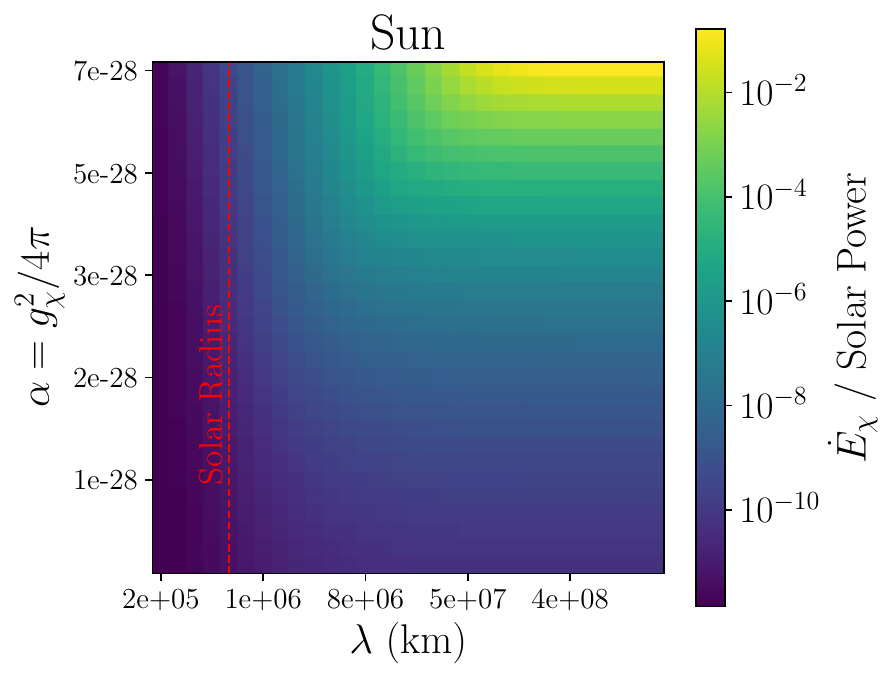}
\includegraphics[width=0.495\linewidth]{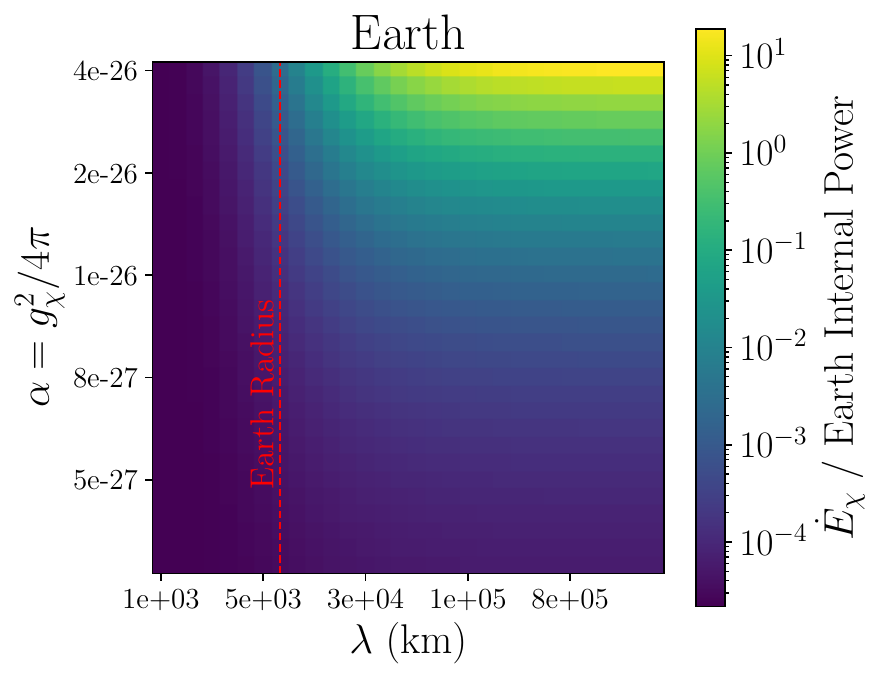}
    \caption{Dark kinetic heating of a benchmark brown dwarf, white dwarf, the Sun, and the Earth, all at the local position as a function of both interaction strength $\alpha$ and dark force range $\lambda$ for 1 GeV dark matter. All these plots assume the object's maximum capture rate. This map shows pure dark kinetic heating with no Standard Model background. The red dashed line on each plot corresponds to the radius of the given object. For the Sun and the Earth, we show the dark kinetic energy deposition rate $\dot{E}_\chi$ as a fraction of the object's measured power, see text for details.}
    \label{fig: heatmap}
\end{figure*}

\begin{itemize}
    \item \textit{White dwarfs:} White dwarfs are smaller but more massive than a brown dwarf, and have gravitational capture rates similar to that of a brown dwarf. However, white dwarfs have higher background temperatures, and thus require more dark heating to provide a detectable signal. White dwarfs have previously been considered in the context of dark matter heating~\cite{Mochkovitch:1985vi,Moskalenko:2006mk,Hooper:2010es,McCullough:2010ai,Krall:2017xij,Curtin:2020tkm,Panotopoulos:2020kuo,Bell:2021fye,Ramirez-Quezada:2022uou,DeRocco:2022rze}, but only in Globular clusters where the dark matter content may be high enough to overcome large heat backgrounds (though the dark matter content in considered systems such as Messier 4 is highly uncertain~\cite{Moore:1995pb,Saitoh:2005tt}). Other recent analyses have explored the ignition of Type-Ia supernovae from localized heating by heavy asymmetric point-like dark matter \cite{Bramante:2015cua,Acevedo:2019gre,Janish:2019nkk} and composite states \cite{Graham:2018efk,Acevedo:2020avd,Raj:2023azx}. Thanks to our dark kinetic heating effect, dark matter signals can be possible in the local position where the dark matter density is instead well known. We consider a benchmark white dwarf with parameters equal to the white dwarf WD 0552-002 \cite{Vincent_2023}, which has an effective temperature of about 5000~K. Note that the thermal heating equilibrium timescale for white dwarfs was not previously calculated in the literature to our knowledge, and so we detail it in Appendix~\ref{app:wdtherm}.
    
    \item \textit{Nuclear Burning Stars:} Nuclear burning stars, such as our Sun, also have non-relativistic escape velocities without the additional dark potential. However, because of the more complex nature of how energy deposition corresponds to total temperature in these systems, we will instead compare the kinetic energy deposition rate of dark matter to the total observed power radiated, rather than directly calculate the kinetic heating impact. In principle interesting additional effects could be considered for the Sun or similar stars, but require stellar simulation codes such as \texttt{MESA}~\cite{2011ApJS..192....3P, 2013ApJS..208....4P, 2015ApJS..220...15P, 2018ApJS..234...34P, 2019ApJS..243...10P, 2023ApJS..265...15J} for reliable results. For the parameter space we investigated for brown dwarfs earlier, the solar kinetic heating is a negligible fraction, such that there are no expected observable alterations to the standard solar structure and its evolution, while still providing interesting signals with brown dwarfs or exoplanets.
    
    \item \textit{Earth:} While the Earth is a much less dense target than the others considered thus far, leading to lower potential capture rates, we are able to robustly measure its heat power. Due to its low escape velocity, the Earth was only considered previously in the context of dark matter annihilation heating~\cite{Mack:2007xj,Starkman:1990nj,Bramante:2019fhi} or asymmetric dark matter that collapses into small evaporating black holes \cite{Acevedo:2020gro}. The internal heat power of the Earth has been measured to be $44.2 \pm 1$ TW \cite{Pollack:1993}. As such, we will show the dark energy deposition as a fraction of this value. Note that a long-range dark matter-baryon force can also affect direct detection experiments on Earth, giving complementary signals; this has been previously explored in Refs.~\cite{Davoudiasl:2017pwe,Davoudiasl:2020ypv}.

    \item \textit{Neutron Stars:} Neutron stars pack the mass of the Sun into a tiny $\sim10$~km radius. Their consequent extreme densities provide relativistic escape velocities, making them the only target considered that can generate observable dark kinetic heating with no additional dark escape velocity (see however Ref.~\cite{Gresham:2022biw}, which considered kinetic heating from increased escape velocity on PSR J2144-3933, the coldest neutron star observed to date \cite{Guillot:2019ugf}). Nonetheless, the inclusion of a new long-range self interaction in the dark sector can increase heating signatures. However, unless the dark potential is very close to the dark matter mass, $i.e.$ $|\Phi(R)/m_\chi|\simeq 1$, at which point our calculations may break down, neutron stars are subdominant to other objects.

\end{itemize}

Figure \ref{fig: heatmap} shows the dark kinetic heating as a function of interaction strength $\alpha$ and force range $\lambda$ for our four targets of interest: brown dwarfs/exoplanets, white dwarfs, the Sun, and the Earth. Overall, an important point of this figure is that for all objects, they are optimized to probe force lengths comparable to or larger than their radii, as was expected earlier from our discussion on brown dwarfs. Unless the force range extends outside of the object, the exponential shutoff of the Yukawa force will mean that nearby dark matter particles feel only a weak potential. On the other hand, once the force range is beyond a certain length at a given coupling, no increase in potential is felt, as seen by the saturation  in heating along the top right of each plot in Fig.~\ref{fig: heatmap}. For larger coupling strengths than shown in Fig.~\ref{fig: heatmap}, the dark potential at the surface of each object becomes extreme ($i.e.$ $|\Phi(R)/m_\chi|\simeq 1$), and additional effects other than kinetic heating may be observed; this regime is beyond the scope of this work. Overall, given the dependence of signal strength on force range and coupling, all these objects provide complementary sensitivity to dark sector parameters.

In Fig.~\ref{fig: heatmap}, we have assumed that all the dark matter passing through each object is captured. We first discuss the situation where the scattering cross section is large enough such that all objects satisfy this condition. In this scenario, in order to get enough heating to be comparable to temperatures of the benchmark white dwarf ($\sim 5000$ K), a larger coupling strength $\alpha$ is required compared to brown dwarf signals. While the radius of a white dwarf is smaller than that of a brown dwarf, we also do not get observable heating until roughly the same force range because of the white dwarf's larger background temperature. For the Sun we have shown the dark kinetic energy deposition rate as a fraction of the sun's observed power ($\sim 10^{26}$ Watts). As expected, the Sun is more sensitive to the dark coupling $\alpha$ than other objects by virtue of its size, but only at force ranges larger than the solar radius. On the other hand, Earth is the most sensitive detector for force ranges less than the radius of a brown dwarf.

These qualitative comparisons can change in the scenario that the dark matter-Standard Model scattering cross section is smaller than one or more object's maximum capture cross section. For example, at 1 GeV dark matter mass, the maximum capture cross section for our benchmark brown dwarf, white dwarf, the Sun and the Earth, is roughly $10^{-35}$~cm$^2$, $10^{-42}$~cm$^2$, $10^{-35}$~cm$^2$, $10^{-33}$~cm$^2$ respectively, at the nucleon-level. When comparing multiple objects, if below one object's maximum cross section, but above another's, lowering the cross section while increasing the coupling strength will maintain a fixed heating rate for one while increasing it for the other. For example, in Fig.~\ref{fig: sigma sensitivity}, we have shown sensitivities for our benchmark brown dwarf which are at cross sections below its maximum capture rate, by increasing the coupling $\alpha$, leading to constant heating rates. Comparing with white dwarfs, whose maximum capture cross section is around $10^{-42}$~cm$^2$, these smaller cross sections do not decrease the capture rate away from its maximum. Therefore, for the white dwarf, the increase in the coupling may correspond to an increase in heating rather than a fixed heating rate, depending on what happens as the dark potential at the surface approaches the dark matter mass. This means that at the bottom end of Fig.~\ref{fig: sigma sensitivity}, a white dwarf could potentially be more sensitive for the same combination of parameters. However, white dwarfs are dominantly composed of zero nuclear-spin nuclei, such that their spin-dependent scattering cross section is severely suppressed. On the other hand, a brown dwarf is dominantly hydrogen which has spin-dependent couplings, and therefore the white dwarf would not be competitive for spin-dependent cross sections. Furthermore, in Fig.~\ref{fig: sigma sensitivity}, we have assumed local position dark matter densities, and in the case of inner Galaxy brown dwarf searches, the density is enhanced by orders of magnitude, such that the sensitivity can be much greater than shown here. Brown dwarfs have larger surface areas such that their luminosity for the same temperature is much larger than a white dwarf, making them potentially more detectable in these dense dark matter environments. 

In Fig.~\ref{fig: heatmap}, the fact that there are any parameters for which the Earth is most sensitive is a testament both to the strong effect of our signal regardless of target size, and to the single terawatt precision measurements of Earth's internal power. However, note that at 1 GeV dark matter mass, and the Earth's maximum capture cross section of about $10^{-33}$~cm$^2$, this may be already ruled out by direct detection, depending on the particle physics model. In Fig.~\ref{fig: heatmap} we aim to demonstrate the complementarity in the force range and coupling strength for these objects with different properties; at other dark matter masses these comparisons should still qualitatively hold at cross sections not ruled out by direct detection.

\section{Summary and Outlook}
\label{sec:conclusion}

Celestial bodies are advantageous dark matter detectors, due to their powerful sensitivities to unexplored dark matter parameter space. One signal commonly studied in the literature is the effect of dark matter energy injection on stars and planets. For the highest-escape velocity object, neutron stars, a signal called dark kinetic heating is possible, due to the relativistic escape velocity induced by their extreme densities. Dark matter is sped up to relativistic speeds when entering neutron stars, and if dark matter is successfully captured its deposited kinetic energy alone. For low-escape velocity objects, $i.e.$ every celestial body except for neutron stars, their escape velocities are not relativistic and so any kinetic heating signal was previously thought negligible. Instead, low-escape velocity objects had previously only been considered in the context of dark matter annihilation heating, where the rest-mass energy of dark matter is injected after annihilating inside the object.

We have pointed out and shown that even low-escape velocity objects can actually give rise to dark kinetic heating signals, expanding the discovery potential and constraining power of heating searches with such objects. This can occur if there is a long-range force in the dark sector, which serves to increase the escape velocity of these objects, by creating an additional dark potential which serves to re-inforce gravity. The induced dark escape velocities can be relativistic, speeding up dark matter to deposit large amounts of dark kinetic energy, and no longer requiring annihilation signals for detectable heating.

We demonstrated the effect of this dark escape velocity in the context of exoplanet and brown dwarf searches. This class of dark matter detectors are ideal due to their large surface areas compared to neutron stars, such that they can be much easier to detect. Exoplanets and brown dwarfs are also timely candidates, with the arrival of new and upcoming telescopes such as JWST, Roman, and Rubin well poised to detect many heating candidates in the near future. We showed that dark kinetic heating substantially expands the discovery potential of this search, and can be so large than even local exoplanets and brown dwarfs are afforded striking and easily detectable dark matter heating signals.  Due to the large potential size of the dark kinetic heating signal in exoplanets, we used existing JWST and Wide-field Infrared Survey Explorer data on the local Super-Jupiter WISE 0855-0714, and showed that some dark matter parameter space can indeed already be probed. Further into the center of the Galaxy, dark kinetic heating predicts a sharper probe of the unknown Galactic dark matter distribution, compared to the dark matter annihilation case previously studied. This correlation with the Galactic dark matter density distribution should be discernible with upcoming surveys.

The presence of the dark escape velocity also produces interesting features in the dark matter parameter space. We found that the cross section corresponding to the maximum capture rate flattens around the transition cross section, which was previously reserved only for very high escape velocity objects such as neutron stars and white dwarfs, or other objects in regions with very low dark matter velocities where capture is very efficient. That is, the presence of the dark escape velocity can make capture extremely efficient. We also noted that the dark kinetic heating signal is optimized for force ranges somewhat larger than the radius of the celestial object, such that many celestial objects provide complementary sensitivity to dark sector parameters. As such, we investigated the complementarity of multiple low escape velocity objects not previously considered in the context of dark kinetic heating, such as white dwarfs, nuclear-burning stars such as our Sun, and the Earth. 

Going forward, it would be interesting to investigate the implications of dark kinetic heating in the context of more complete particle physics models. Dark kinetic heating removes the requirement of dark matter annihilation for dark matter heating to be detectable, widening the particle model space for which low-escape velocity objects are relevant. In this work we also did not consider the additional effects of dark matter heating from annihilation, which would be expected from WIMP-like dark matter models, where both dark kinetic heating and annihilation can be relevant. This could increase the discovery or constraining power of these searches, depending on the model setup. On the observational side, there are a plethora of candidates which could already be used in complementary ways to probe particle physics of the dark sector. We showed sensitivity with the Super-Jupiter WISE 0855-0714, but with the increasing power of new telescope technologies, the span of excellent targets will only continue to give us new insights to physics of the dark sector.

\acknowledgements 

We thank Zach Bogorad, Joseph Bramante, Peter Denton, Vincent Lee, Nirmal Raj, Philip Schuster, Juri Smirnov, and Natalia Toro for helpful comments and discussions. JFA, RKL, and AJR are supported by the U.S. Department of Energy under Contract DE-AC02-76SF00515. AJR is also supported by the NSF GRFP under grant DGE-2146755.

\clearpage
\newpage
\onecolumngrid
\appendix

\section{Self-Interaction Jeans Length}
\label{app: Jeans}

We show that the self-interaction considered does not lead to Jeans instabilities on scales of order the interaction range $\lambda$ for halo dark matter. We begin by estimating the free-fall time of a dark matter sphere under this new force, which is given by 
\begin{equation}
    t_{\rm ff} \simeq \sqrt{\frac{3 \pi m_\chi^2}{32 \alpha_\chi \rho_\chi}} \simeq 1.3 \times 10^9 \ {\rm s} \, \left(\frac{\alpha_\chi}{10^{-26}}\right)^{-1/2} \left(\frac{\rho_\chi}{\rm GeV \ cm^{-3}}\right)^{-1/2} \left(\frac{m_\chi}{\rm GeV}\right)~.
\end{equation}
This timescale must be compared to the sound crossing time over the same region; if the sound waves take longer than $t_{\rm ff}$ to cross a region of size $\lambda$, they will be unable to stabilize the clump against collapse via the attractive dark force. Given the extremely weak couplings we analyze, the sound speed of the dark matter $c_s$ can be approximated as that of an ideal gas  
\begin{equation}
    c_s \simeq v_p \simeq 270 \ \rm km \ s^{-1}\,,
\end{equation}
and, therefore, the sound crossing time across a region of size $\lambda$ is approximately
\begin{equation}
    \left(\frac{c_s}{\lambda}\right)^{-1} \simeq 2.5 \times 10^5 \ {\rm s} \, \left(\frac{\lambda}{100 R_\odot}\right) \left(\frac{v_p}{270 \ \rm km \ s^{-1}}\right)^{-1}\,,
\end{equation}
which is always much shorter than the free-fall time for $\lambda \lesssim 100 \, R_\odot$. Therefore, across the range of couplings, dark matter masses, and mediator ranges considered, we do not expect the new dark force to induce a Jeans instability in the Galactic halo dark matter.  More broadly, new effects other than dark kinetic heating may be observable, and would be interesting to consider as new dark sector discovery avenues, but are outside the scope of this work.

\section{$\Phi$ Screening in High Dark Matter Densities}
\label{app: screening}

We now analyze corrections to the effective potential given by Eq.~\eqref{eq:Phi_gauss}. In general, for a finite-range Yukawa potential, the field outside a dark matter distribution with some characteristic radius $R_\chi$ will be given by
\begin{equation}
    \label{eq: app Phi full}
    \Phi(r \geq R_\chi) = \Phi(R_\chi) \left(\frac{R_\chi}{r}\right) e^{-m_\phi(r - R_\chi)}~,
\end{equation}
where $\Phi(R_\chi)$ is obtained by solving the field equation both inside and outside the dark matter distribution, and imposing continuity and differentiability at the boundary $r = R_\chi$. In the regime where $m_\phi R_\chi \ll 1$, this procedure yields 
\begin{equation}
\label{eq: app Phi Gauss}
    \Phi(R_\chi) = - \frac{N_\chi\alpha}{R_\chi} e^{-m_\phi R_\chi} ~,
\end{equation}
and thus Eq.~\eqref{eq:Phi_gauss} is recovered. In other words, the field is effectively Coulomb-like in this limit.

This solution is altered in the presence of a quartic $y \phi^4$ interaction. While we have we have assumed $y = 0$ at tree-level in Eq.~\eqref{eq: lagrangian}, a quartic term with coupling strength $y \propto g_\chi^4$ will be generically induced at the loop-level by a box diagram with $\chi$ running in the loop. In principle a cubic interaction is also allowed and will have a similar effect; for simplicity we do not consider this, though note that it is technically natural to set such a term to zero.
A dark matter background density will then generate a vacuum expectation value $\langle \phi \rangle$ which shifts the mass of the field $\phi$ to some effective value $m_{\phi}^{*} > m_\phi$ within $R_\chi$~\cite{Blinov:2018vgc, Denton:2023iaa}. If the new force range $\lambda^{*} = 1/m_\phi^*$ is small enough such that $\lambda^* \lesssim R_\chi$, then Gauss' law no longer applies and instead we must solve for $\Phi(R_\chi)$ more generally, which will no longer be given by Eq.~\eqref{eq: app Phi Gauss}. When $\lambda^* \gg R_\chi$, Eq.~\eqref{eq: app Phi full} simply reduces to Eq.~\eqref{eq: app Phi Gauss}. To calculate the size of $m_{\phi}^*$, recall the Lagrangian for $\phi$, 
\begin{equation}
\label{eq: app phi lagrangian}
   \mathcal{L} \supset \frac{1}{2}(\partial_\mu \phi)^2 - \frac{1}{2}m_\phi^2 \phi^2 - g_\chi \phi \bar{\chi} \chi + \frac{1}{4!}y\phi^4.
\end{equation}
Furthermore, in the presence of a large dark matter number density $n_\chi$,
\begin{equation}
    g_\chi \phi \bar{\chi}\chi \rightarrow g_\chi \phi n_\chi.
\end{equation}
This leads to a non-zero vacuum expectation value for $\phi$ set by minimizing the potential,
\begin{equation}
V(\phi) = \frac{1}{2}m_\phi^2 \phi^2 + g_\chi \phi n_\chi - \frac{1}{4!}y\phi^4.
\end{equation}
The exact form of the vacuum expectation value is not elucidating, but expanding around $\langle \phi \rangle$ gives us
\begin{equation}
    m_{\phi}^* = \sqrt{m_\phi^2 + \frac{1}{2}y\langle\phi\rangle^2}.
\end{equation}
We therefore require that for all times, $\lambda^* \gg R_\chi$ . Eventually a celestial object may accumulate enough dark matter such that this condition is not satisfied, and heating will stop increasing as rapidly. For some of the values of $m_\phi$ and $g_\chi$ considered here, this requires $y \ll g_\chi^4$. However, this occurrence is very strongly dependent on dark matter mass, dark matter density, $g_\chi$, and the object in question. For example, at $m_\chi \lesssim 10$ MeV, significant heating occurs at such low values of $g_\chi$ that setting $y \simeq g_\chi^4$ does not qualitatively change our results for the Earth. For other objects, such as brown dwarfs, a similar statement can be made for reasonable non-local dark matter densities.

\section{Maximum Impact Parameter Calculation}
\label{app: impact param}

We calculate the maximum impact parameter that a passing dark matter particle can have and still be pulled in to pass through the surface of the celestial body in question. We assume a Schwarzchild metric in the space near the body. We start with the equation for radial velocity, Eq.~\eqref{eq: radial motion scalar mediator}, which we reiterate here
\begin{equation}
    \begin{split}
        \label{app eq: radial motion scalar mediator}
        \left(\frac{dr}{d\tau}\right)^2   = \frac{\mathcal{E}^2}{(m_\chi+\Phi)^2} - \left(1-\frac{2GM}{r}\right)\left(1+\frac{L^2}{r^2(m_\chi+\Phi)^2}\right)~,
    \end{split}
\end{equation}
\begin{equation}
    \begin{split}
    \mathcal{E} = m_\chi\gamma_\chi~, \\
    L = b \, m_\chi \gamma_\chi  v_\chi~,
\end{split}
\end{equation}
where $\mathcal{E}$ and $L$ are the dark matter particle's energy (including the gravitational binding energy due to gravity) and orbital angular momentum respectively. To find the maximum impact parameter for which a dark matter particle will hit the object, we will separate out two cases. We first set the point at which a centrifugal barrier is reached ($dr/d\tau = 0$) to be at the radius of the object $R$. Setting Eq.~\eqref{app eq: radial motion scalar mediator} to zero and solving for $b$ we find
\begin{equation}
    \label{eq: bmax scalar mediator}
    b_{\rm max}^{\rm dark} = \frac{R}{v_\chi}\sqrt{\frac{v_\chi^2 + \frac{2GM}{R\gamma_\chi^2} - \frac{2\Phi}{m_\chi\gamma_\chi^2} - \frac{\Phi^2}{m_\chi^2\gamma_\chi^2} + \frac{2GM}{R}\left(\frac{2\Phi}{m_\chi\gamma_\chi^2} + \frac{\Phi^2}{m_\chi^2\gamma_\chi^2}\right)}{1-\frac{2GM}{R}}}~,
\end{equation}
For reasons we detail in Appendix~\ref{app: velocity boost scalar potential}, we will stay within the regime where $|\Phi(r > R)| \lesssim m_\chi$. In particular, for $|\Phi| \ll m_\chi$, we can drop terms of $\mathcal{O}(|\frac{\Phi}{m_\chi}|^2)$ as well as terms $\propto G\frac{\Phi}{m_\chi}$, which are small. Finally, approximating also $\gamma_\chi \simeq 1$, since the dark matter particle is non-relativistic far from the celestial body, we obtain Eq.~\eqref{eq: bmax exp} of the main text,
 \begin{equation}
 \label{eq: app bmax approx}
    b_{\rm max}^{\rm dark} \simeq R \times \sqrt{\frac{1 + \frac{2GM}{R v^2_\chi} - \frac{2\Phi}{m_\chi v^2_\chi}}{1-\frac{2GM}{R}}}~,
\end{equation}
which corresponds to the exponential growth phase of the capture rate discussed in the main text.

Once enough dark matter has been accumulated, we can no longer drop terms and must use the full form of Eq. ~\eqref{eq: bmax scalar mediator}. Furthermore, we have found so far the impact parameter such that there is a centrifugal barrier at $R$. However, this does not preclude the possibility of a second centrifugal barrier at a position that lies beyond $R$. This second barrier is related to the exponential decay of the long-range force as described in Section \ref{accumulation}, and whose existence can be seen from the fact that there can in principle be multiple zeros of Eq.~\eqref{app eq: radial motion scalar mediator}. The existence of this second barrier while using $b_{\rm max}^{\rm dark}$ would lead to overestimates of the dark matter flux. 

When there is a secondary centrifugal barrier, it will be due to a local maximum in the term
\begin{equation}
    \begin{split}
        V_{\rm eff}^2 = \left(1-\frac{2GM}{r}\right)\left((m_\chi+\Phi)^2+\frac{L^2}{r^2}\right)~,
        \label{eq:app_V_eff}
    \end{split}
\end{equation}
which makes $dr/d\tau$ vanish, $cf.$ Eq.~\eqref{app eq: radial motion scalar mediator}. Since Eq.~\eqref{eq:app_V_eff} determines the centrifugal barrier(s), we refer to this quantity as the (squared) effective potential, in analogy with the pure Newtonian case. Centrifugal barriers exist when $V_{\rm eff}^2 = \mathcal{E}^2$, and a second barrier beyond the one we require to be at $R$ may appear due to additional local maxima of $V_{\rm eff}$. We thus define the ``true" maximum impact parameter as 
\begin{equation}
b_{\rm max} = \min\left(b_{\rm max}^{\rm dark}, \tilde{b}_{\rm max}^{\rm dark}\right), 
\end{equation}
where $\tilde{b}_{\rm max}^{\rm dark}$ is as large as possible such that $V_{\rm eff}^2 \leq \mathcal{E}^2$ for all $r > R$. This procedure minimizes $b_{\rm max}$ until there exists no centrifugal barrier at radii larger than R. To find $\tilde{b}_{\rm max}^{\rm dark}$, we must solve for the parameters that yield $V_{\rm eff}^2 = \mathcal{E}^2$ and also produce a local maximum $V_{\rm eff}' = 0$. This condition imposes that the second centrifugal barrier is not enough to block in-falling particles. To simplify the problem, we approximate $1 - 2GM/r \simeq 1$ for $r>R$, which is reasonable for all the celestial bodies we consider. This results in the following system of equations
\begin{equation}
    \begin{split}
    \label{eq: outer barrier conditions}
    {\rm (I)} \quad & (m_\chi + \Phi)^2 + \frac{L^2}{r^2} = \mathcal{E}^2 ~, \\
    {\rm (II)} \quad & (m_\chi+\Phi)\Phi' - \frac{L^2}{r^3} = 0  ~.
    \end{split}
\end{equation}
This system must be solved numerically. To do so, we first analytically isolate $b$ from $\rm (I)$ in Eq.~\eqref{eq: outer barrier conditions}, since isolating $r$ is more difficult due to the mixed, exponential-polynomial dependence through $\Phi$. We then replace $b$ in $\rm (II)$ by plugging it into $L$, and numerically find the value of $r$ that solves the equation. This second barrier only exists for relatively large values of $\alpha/\lambda$, where the exponential decay of the Yukawa potential at $r>\lambda$ becomes the limiting factor, so Eq.~\eqref{eq: app bmax approx} is typically the relevant equation for most of our parameter space of interest. Note that when $b_{\rm max} = \tilde{b}_{\rm max}^{\rm dark}$ dark matter accumulation is no longer exponential but rather sub-linear.

\section{Velocity Boost with Scalar Potential}
\label{app: velocity boost scalar potential}

\subsection{Velocity Boost Calculation}
We compute the boost of a dark matter particle due to the new long-range potential $\Phi$. In our calculation, we will assume energy is conserved during the initial approach of the dark matter to the celestial body. We analyze the validity of this assumption below, and show it is a good approximation. We start from the action of the particle in flat spacetime, which reads 
\begin{equation}
    \begin{split}
        S = - \int (m_\chi + \Phi) \, d\tau~, 
    \end{split}
\end{equation}
where $\tau$ is proper time in the particle's rest frame. The equation of motion for the time component is
\begin{equation}
\label{eq: dt/dtau minkowski}
        \frac{d}{d\tau} \left((m_\chi + \Phi) \,  \frac{dt}{d\tau}\right) = g_\chi \frac{\partial\phi}{\partial \tau} = 0 ~.\\
\end{equation}
Integrating this and using 
\begin{equation}\frac{dt}{d\tau} = \frac{1}{\sqrt{1-v(r)^2}}~,
\end{equation}
where $v$ is the 3-velocity modulus, yields
\begin{equation}
    \frac{m_\chi + \Phi}{\sqrt{1-v^2(r)}} = \mathcal{E} ~,
\end{equation}
where the constant of motion $\mathcal{E} > 0$ is the particle's energy, and is fixed further below from the initial condition. From the above equation, we can solve for the velocity 
\begin{equation}
    v^2(r) = 1 - \left(\frac{m_\chi}{\mathcal{E}}\right)^2 \left(1 + \frac{\Phi(r)}{m_\chi}\right)^2~.
    \label{eq:boost_sr}
\end{equation}
This indicates that there exists a finite value of $\Phi = -m_\chi$ for which the particle seemingly reaches the speed of light. This is of course non-physical and as $|\Phi| \rightarrow m_\chi$, other physical effects become significant that are outside the scope of this work. We avoid the parameter space where the scalar potential reaches this critical value at the surface of the object throughout its evolution.

Incorporating the effect of a gravitational field into the previous calculation is straightforward. The integrated equation of motion is modified by an additional metric component factor \cite{landau2013classical},
\begin{equation}
\label{eq: dt/dtau schwarzschild}
    (m_\chi+\Phi)g_{00}\frac{dt}{d\tau} = \mathcal{E}~,
\end{equation}
which outside of the celestial body evaluates to the Schwarzchild solution
\begin{equation}
\label{eq: g_00}
    g_{00}(r) = 1- \frac{2GM}{r}~.
\end{equation}
We then write 
\begin{equation}
d\tau = \sqrt{g_{00}(r)dt^2 - dz^2}~,
\end{equation}
where 
\begin{equation}
    dz = \sqrt{g_{00}^{-1}dr^2 + r^2d\Omega^2}~,
\end{equation} is the infinitesimal proper distance measured by a local observer, with $d\Omega$ the solid angle. Defining the velocity as 
\begin{equation}
    v = \frac{dz}{\sqrt{g_{00}} dt}
\end{equation}
we then have 
\begin{equation}
    \mathcal{E} = \frac{\left(m_\chi+\Phi\right) \sqrt{g_{00}}}{\sqrt{1 - v^2}} = \left(m_\chi+\Phi\right) \sqrt{g_{00}} \, \gamma_r~,
\end{equation}
where $\gamma_r$ is the corresponding boost factor at the coordinate $r$. Solving for this, we obtain
\begin{equation}
    \gamma_r = \frac{\mathcal{E}}{\left(m_\chi+\Phi\right)\sqrt{g_{00}}} = \gamma_\chi\left[\left(1 + \frac{\Phi(r)}{m_\chi}\right) \sqrt{1-\frac{2GM}{r}}\;\right]^{-1} ~,
\end{equation}
where we have fixed the energy from the initial condition to be $m_\chi \gamma_\chi$. Because the dark matter is initially non-relativistic far from the celestial body, its energy will be approximately
\begin{equation}
   \mathcal{E} = \frac{m_\chi}{\sqrt{1-v_\chi^2}} \simeq m_\chi + \frac{1}{2} m_\chi v_\chi^2~,
\end{equation}
where the typical halo velocity is of order $v_\chi \sim v_p \simeq 10^{-3}$. 

\subsection{Time Variation of Energy}
Strictly speaking, the dark matter's energy does not remain constant prior to being captured as it approaches the celestial object. Firstly, the potential felt by the particles is steadily increasing as more dark matter is captured over time. Secondly, the in-falling dark matter particles can radiate the light mediator as they accelerate. In the latter case, the energy loss rate would scale as $\propto \alpha^3$, and therefore is extremely suppressed in our parameter space of interest. We now address the former case, and demonstrate that the time variation of the potential is negligible during the average time it takes for a dark matter particle to be captured. This holds true even in the regime where dark matter accumulation, and therefore the long-range potential, grows exponentially. Hence, our calculation of the boost above assuming energy conservation is highly accurate.

Denoting the timescale for a dark matter particle to fall onto the celestial object as $t_{\rm fall}$, the condition
\begin{equation}
    \int_{t}^{t + t_{\rm fall}} \frac{\dot{\Phi}(t')}{\Phi(t')} \, dt' \ll 1
    \label{eq:E_const_cond_1}
\end{equation}
must be met at any given time $t$ in order to neglect the variation of the potential during the dark matter capture process. In other words, the potential must not vary significantly during the falling timescale. To be maximally conservative, we will assume the non-screened case where the dark potential is proportional to the number of dark matter particles, $i.e.$ $\Phi \propto N_\chi$. We also consider the exponential growth phase, since in this regime the potential varies the fastest over time. This means that
\begin{equation}
    \frac{\dot{\Phi}}{\Phi} = \frac{\dot{N}_\chi}{N_\chi} = \frac{c_0 + c_1 N_\chi}{N_\chi} \simeq c_1~,
\end{equation}
where we have used Eq.~\eqref{eq:Nchi_diff}, and assumed $N_\chi \gg 1$ on the rightmost side. Thus, Eq.~\eqref{eq:E_const_cond_1} is reduced to
\begin{equation}
        c_1 \times t_{\rm fall} \simeq 10^{-17} \, \left(\frac{\alpha}{10^{-27}}\right) \left(\frac{R}{R_{\rm Jupiter}}\right) \left(\frac{\rho_\chi}{0.4 \ \rm GeV \ cm^{-3}}\right) \left(\frac{m_\chi}{\rm GeV}\right)^{-2} \left(\frac{v_\chi}{270 \ \rm km \ s^{-1}}\right)^{-1} \left(\frac{t_{\rm fall}}{\rm s}\right)~.
\end{equation}
To derive the above scaling, we have assumed $\lambda \gg R$ as well as a capture probability $P_{\rm cap} = 1$, $cf.$ Eq.~\eqref{eq: c1}. This choice maximizes the rate of growth of the potential, and so our estimate for $c_1$ is highly conservative for our purposes here. The above equation must then be much less than one for the constant energy approximation to be reasonable. 

As an extremely conservative upper bound on the falling time, we neglect the dark matter potential and use the gravitational free-fall time. In doing so, we assume a particle that is at rest at an initial distance $r_{\rm init} \sim 10 \lambda$, which is about the maximum distance at which dark matter particles are typically captured in our setup. The resulting time is
\begin{equation}
    t_{\rm fall}^{\rm grav} = \frac{\pi r_{\rm init}^{3/2}}{\sqrt{2 G M}} \simeq 1.5 \times 10^7 \ {\rm s} \left(\frac{\lambda}{100 \, R_{\rm Jupiter}}\right)^{\frac{3}{2}} \left(\frac{M}{55 \, M_{\rm Jupiter}}\right)^{-\frac{1}{2}}~.
\end{equation}
We see that we have $c_1 t_{\rm fall} \ll 1$. This implies that the time variation of the dark potential $\Phi$ can be completely neglected when considering the evolution of an individual dark matter particle as it approaches the celestial body. We emphasize again that this estimate is extremely conservative as we have considered the exponential-growth phase of the dark potential while we have only used the purely-gravitational free-fall time. Accounting for the full effect of the dark potential would only serve to speed up the dark matter further and shorten the falling timescale.

\section{Calculating $\langle{(b_{\rm max}^{\rm dark})^2 v_\chi \rangle}$}
\label{app: Ndot}

We provide further details on how Eq.~\eqref{eq: Ndot approx} in the main text is obtained. We can evaluate the expression $\langle (b_{\rm max}^{\rm dark})^2 v_\chi P_{\rm cap}\rangle$ by treating $b_{\rm max}^{\rm dark} \, v_\chi$ approximately as a constant,
\begin{equation}
    \langle (b_{\rm max}^{\rm dark})^2 v_\chi P_{\rm cap}\rangle \simeq (b_{\rm max}^{\rm dark} v_\chi)^2 \langle\frac{P_{\rm cap}}{v_\chi} \rangle = b_{\rm max}'^2 \langle\frac{P_{\rm cap}}{v_\chi} \rangle~,
\end{equation}
where 
\begin{equation}
    b_{\rm max}' = R\times\sqrt{\frac{\frac{2GM}{R } - \frac{2\Phi}{m_\chi }}{1-\frac{2GM}{R}}}~.
\end{equation}
From Eq.~\eqref{eq: bmax exp} we can see that this is valid up to terms of order $v_\chi$. For the general case we evaluate $\langle P_\text{cap}/v_\chi \rangle$ numerically, but for sufficiently high cross sections we can simply set $P_{\rm cap} = 1$.  We then simply integrate against a Maxwell-Boltzmann distribution to find 
\begin{equation}
    \begin{split}
    \langle (b_{\rm max}^{\rm dark})^2 v_\chi\rangle & \simeq b_{\rm max}'^2 \frac{4}{\sqrt{\pi}v_p^3}\int_0^{v_{\rm max}} v_\chi \, e^{-v_\chi^2/v_p^2} \, dv_\chi \\
    & = b_{\rm max}'^2 \frac{2}{\sqrt{\pi}v_p}(1-e^{-v_{\rm max}^2/v_p^2}) \\
    & \simeq b_{\rm max}'^2 v_p~,
    \end{split}
\end{equation}
where $v_p$ is the most probable dark matter speed, and $v_{\rm max}$ is the escape speed of the galaxy which ranges $\sim 500 - 1000 \ \rm km \ s^{-1}$ between our local position and the Galactic Center \cite{2013A&A...549A.137I}.

\section{Initial Dark Matter Growth versus Evaporation}
\label{app:dm_evap}

In Section~\ref{sec: cross section} we noted that in some parts of the parameter space, evaporation could potentially prevent the initial buildup of dark matter, but that this is highly model-dependent \cite{Acevedo:2023owd}. We expand upon this point, and show as a simple example how evaporation can be mitigated at all times, including when very little dark matter has been accumulated, with a minimal extension of our setup whereby $\phi$ is also weakly-coupled to the Standard Model through a term $g_n \phi \bar{n} n$. For brevity, we will only consider our benchmark brown dwarf, although the analysis can be straightforwardly extended to other celestial bodies. 

The rate at which dark matter evaporates away is exponentially dependent on the ratio between temperature and escape energy, where the latter is determined by the external forces acting on the dark matter. Thus, even a slight increase in the escape energy will significantly suppress the evaporation rate, consequently reducing the evaporation mass. When dark matter experiences only gravity without additional external forces, the escape energy equals the gravitational potential energy. Evaluating this quantity at the evaporation mass indicates the threshold at which evaporation would become significant, since lighter particles would evaporate away. As a simple order-of-magnitude estimate, at the center of a celestial body this is 
\begin{equation}
    |\Phi_{\rm grav}(r = 0)| \sim \frac{G M m_\chi^{\rm evap}}{R} \simeq 5 \times 10^{-7} \ {\rm GeV} \left(\frac{M}{55 M_{\rm Jupiter}}\right) \left(\frac{m_\chi^{\rm evap}}{0.5 \ \rm GeV}\right) \left(\frac{R}{R_{\rm Jupiter}}\right)^{-1}~,
    \label{eq:app_phi_evap_1}
\end{equation} 
where in the rightmost expression we have taken a fiducial evaporation mass of $500 \ \rm MeV$, corresponding to a cross-section of about $\sigma_{\chi n} \simeq 10^{-32} \ \rm cm^2$ for non-annihilating dark matter in a $55 M_{\rm Jupiter}$ brown dwarf \cite{Leane:2022hkk}. We compare this quantity to $|\Phi(r = 0)|$ which, for a range $\lambda \gtrsim R$, will have a similar scaling of the form 
\begin{equation}
    |\Phi(r = 0)| \sim \frac{g_n g_{\chi} N_{\rm SM}}{R} \simeq 3 \times 10^{-7} \ {\rm GeV}\left(\frac{g_n}{10^{-24}}\right) \left(\frac{\alpha}{10^{-27}}\right)^{\frac{1}{2}} \left(\frac{M}{55 M_{\rm Jupiter}}\right) \left(\frac{R}{R_{\rm Jupiter}}\right)^{-1}~,
    \label{eq:app_phi_evap_2}
\end{equation}
where we have normalized $g_n = 10^{-24}$ in the rightmost expression. This is about the largest coupling allowed by current fifth force searches \cite{Schlamminger:2007ht,Fayet:2017pdp} for forces of astrophysical range. Note that this does not account for potential screening effects, which might relax such constraints \cite{Blinov:2018vgc}. 

Comparing Eq.~\eqref{eq:app_phi_evap_1} to Eq.~\eqref{eq:app_phi_evap_2} we see that, for the range of dark matter self-couplings we have considered, the long-range potential sourced by the celestial body would significantly increase the total escape energy. This in turn curbs the evaporation rate substantially, and therefore reduces the evaporation mass compared to the naive calculation that only assumes gravity and temperature play a role. 

 Above, we have only briefly illustrated one possible scenario by which the evaporation mass can be substantially reduced. More broadly, this simple example estimate does not include self-interaction effects on evaporation, which may also significantly suppress evaporation in our case. Furthermore note that we have not included the fact that the object once heated by dark matter can increase the temperature relevant for evaporation; however we expect this effect to be small given our large escape velocities.

\section{White Dwarf Thermal Equilibrium Timescales}
\label{app:wdtherm}

In degenerate objects, quantum effects become relevant and the heat capacity deviates from the perfect gas approximation we made above in Sec.~\ref{sec:therm} for planets and brown dwarfs. The timescale for equilibration between dark matter heating and cooling in degenerate objects such as neutron stars has been explored in prior works. For completeness, we estimate here the timescale for another degenerate object, a typical white dwarf, to reach an equilibrium temperature for dark matter heating. 

For white dwarfs, there are two relevant temperature scales to consider: the temperature at which a body-centered-cubic ion lattice forms in a first-order phase transition \cite{1975ApJ...200..306L,PhysRevE.103.043204}, 
\begin{equation}
    T_g \simeq 3 \times 10^8 \ {\rm K} \, \left(\frac{\rho}{ 10^7 \ \rm g \ cm^{-3}}\right)^{1/3}~,
\end{equation} 
and the Debye temperature $T_D$ at which quantum effects become relevant, which is approximately given by the ion plasma frequency $\omega_p$  \cite{Shapiro:1983du},
\begin{equation}
    T_D \simeq \omega_p = \sqrt{\frac{4 \pi Z^2 e^2 \rho}{m^2_N}} \simeq 4.4 \times 10^7 \ {\rm K} \, \left(\frac{\rho}{10^7 \ \rm g \ cm^{-3}}\right)^{\frac{1}{2}} \left(\frac{Z}{6}\right) \left(\frac{A}{12}\right)^{-1}~,
\end{equation}
where $\rho$ is the mass density, $m_N = A m_n$ is the ion mass with $m_n$ the nucleon mass, and $Z$ and $A$ are the atomic and mass number of the ions, respectively. For most white dwarfs, $T_D \lesssim T_g$ throughout almost all the volume. 

The white dwarf ions' heat capacity varies between the ranges defined by the above temperatures. 
At temperatures $T \gtrsim T_g$, the white dwarf plasma may be reasonably modeled as a weakly-coupled monoatomic gas, with a heat capacity given by Eq.~\eqref{eq:Cv_main} with $\Gamma = 5/3$. At temperatures $T_D \lesssim T \lesssim T_g$, ions vibrate around their lattice positions but quantum effects are still largely irrelevant. The heat capacity remains temperature-independent but is now increased because ion-ion interactions become important. Using the equipartition theorem, valid in the classical regime $T \gtrsim T_D$, one obtains $C_v \simeq 3N_{\rm SM}$. Finally, at temperatures $T \lesssim T_D$, the heat capacity is dominated by the phonon spectrum of the lattice, and scales as \cite{Shapiro:1983du}
\begin{equation}
   C_v \simeq \frac{16 \pi^4}{5} \left(\frac{T}{T_D}\right)^3~.
\end{equation}

White dwarfs in late evolution stages predominantly cool through photon emission from the thin non-degenerate layer near the surface. We use a simple estimate for the cooling rate as obtained from Ref.~\cite{Shapiro:1983du}, which integrates the luminosity assuming a diffusive regime for photons up to the radius at which the pressure is no longer dominated by electron degeneracy. For a typical carbon white dwarf, this procedure yields a total cooling rate
\begin{equation}
    \dot{E}_{\rm cool} \simeq 4 \times 10^{33} \ {\rm GeV \ s^{-1}} \left(\frac{M}{M_\odot}\right) \left(\frac{T_{c}}{10^7 \ \rm K}\right)^{3.5}~,
    \label{eq:wd_cool}
\end{equation}
where $T_c$ is the temperature of the white dwarf's core, which is approximately isothermal. We use this cooling rate, combined with the heat capacity of the ion lattice, to estimate the thermal equilibration timescale. Eq.~\eqref{eq:therm_evol_main} is integrated as before, but with the replacement $T \rightarrow T_c$. Once we obtain the time evolution of the core temperature through this procedure, we can calculate the effective temperature of the white dwarf. By definition, this is determined as the temperature at which a perfect blackbody of the same size as the white dwarf would radiate at the same rate as that given by Eq.~\eqref{eq:wd_cool}.

\begin{figure}[t]
\centering
    \includegraphics[width=0.5\linewidth]{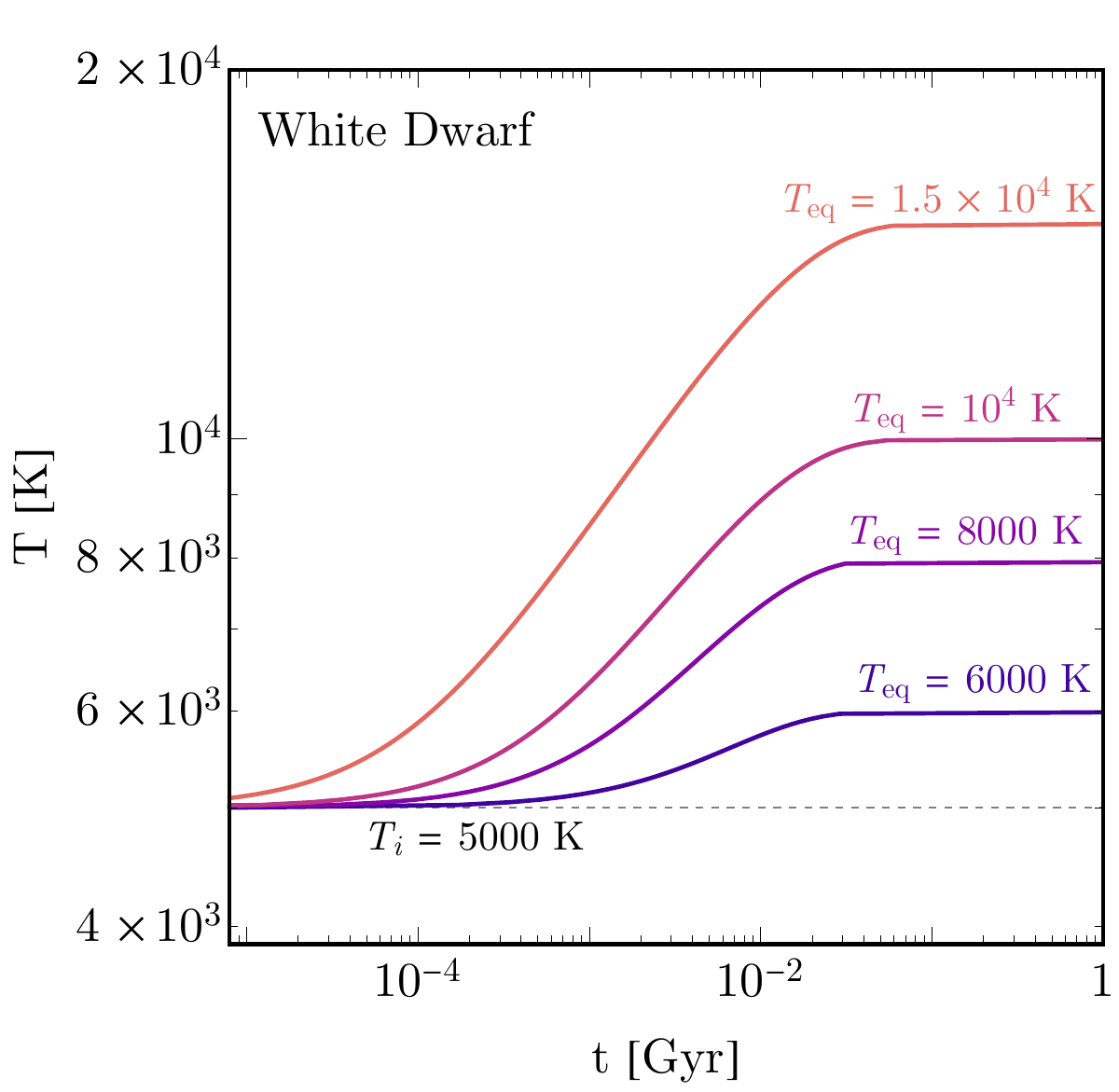}
\caption{Effective temperature evolution for our benchmark white dwarf, assuming fixed energy injection rates. The dashed line indicates the initial effective temperature assumed. Each colored label shows the final equilibrium effective temperature $T_{\rm eq}$ reached.}
\label{fig:WD_heat_timescale}
\end{figure}

Figure~\ref{fig:WD_heat_timescale} shows the evolution of the effective temperature of a solar-mass pure carbon white dwarf for various fixed energy injection rates, initially starting from $5000 \ \rm K$ as per our benchmark white dwarf described in Section~\ref{sec: other objects}. Note that this initial effective temperature corresponds to a core temperature of about $T_c \simeq 5.5 \times 10^6 \ \rm K$ which lies well within the quantum regime described above. In all cases, it can be seen that the timescale to reach thermal equilibrium is much shorter than the typical lifetime of these objects, of order multi-gigayear. We also note that, because the heat capacity increases with temperature in this temperature range, the timescale to reach thermal equilibrium is longer for a larger heating rate. This is in contrast with the fixed heat capacity case we showed for planets and brown dwarfs, where a larger heating rate also led to a shorter timescale to reach equilibrium.

\bibliographystyle{apsrev4-1} 
\bibliography{refs}

\end{document}